\documentclass{jfm}

\usepackage{amssymb,amsmath}
\usepackage{epsfig}
\usepackage{bm}
\usepackage{color}
\usepackage{epstopdf}
\usepackage{graphicx}
\usepackage{natbib}

\newcommand\beq{\begin{equation}}
\newcommand\eeq{\end{equation}}
\newcommand\beqa{\begin{eqnarray}}
\newcommand\eeqa{\end{eqnarray}}

\newcommand{\al}{\alpha}

\title[Kinetic theory of granular particles]{Kinetic theory of granular particles immersed in a molecular gas}

\author[Rub\'en G\'omez Gonz\'alez and Vicente Garz\'o]%
{R\ls U\ls B\ls \'E\ls N\ls \ns G\ls \'O\ls M\ls E\ls Z G\ls O\ls N\ls Z\ls \'A\ls L\ls E\ls Z\ls$^1$ and
V\ls I\ls C\ls E\ls N\ls T\ls E\ls \ns G\ls A\ls R\ls Z\ls \'O\ls$^2$}

\affiliation{$^1$Departamento de F\'isica, Universidad de Extremadura, Avda. de Elvas s/n, 06006  Badajoz (Spain)
\\[\affilskip]
$^2$ Departamento de F\'isica, Instituto Universitario de Computaci\'on Cient\'ifica Avanzada (ICCAEx), Universidad de Extremadura, Avda. de Elvas s/n, 06006  Badajoz (Spain).}

\begin{document}

\maketitle

\begin{abstract}
The transport coefficients of a dilute gas of inelastic hard spheres immersed in a molecular gas are determined. We assume that the number density of the granular gas is much smaller than that of the surrounding molecular gas, so that the latter is not affected by the presence of solid particles. In this situation, the molecular gas may be treated as a thermostat (or bath) of elastic hard spheres at a fixed temperature. This system (granular gas thermostated by a bath of elastic hard spheres) can be considered as a reliable model for describing the dynamic properties of particle-laden suspensions. The Boltzmann kinetic equation is the starting point of the present work. First step is to characterise the reference state in the perturbation scheme, namely the homogeneous state. Theoretical results for the \textit{granular} temperature and kurtosis obtained in the homogeneous steady state are compared against Monte Carlo simulations showing a good agreement. Then, the Chapman--Enskog method is employed to solve the Boltzmann equation to first order in spatial gradients. As expected, the Navier--Stokes--Fourier transport coefficients of the granular gas are given in terms of the solutions of a coupled set of linear integral equations which are approximately solved by considering the leading terms in a Sonine polynomial expansion. Our results show that the dependence of the transport coefficients on the coefficient of restitution is quite different from that found when the influence of the interstitial molecular gas is neglected (dry granular gas). When the granular particles are much more heavier than the gas particles (Brownian limit) the expressions of the transport coefficients are consistent with those previously derived from the Fokker--Planck equation. Finally, as an application of the theory, a linear stability analysis of the homogeneous steady state is performed showing this state is always linearly stable.
\end{abstract}

\date{\today}
\maketitle

\section{Introduction}
\label{sec1}

A challenging problem in statistical physics is the understanding of multiphase flows, namely, the flow of solid particles in two or more thermodynamic phases. Needless to say, these type of flows occur in many industrial settings (such as circulating fluidised beds) and can also affect our daily lives due to the fact that the comprehension of them may ensure vital needs of humans such as clean air and water \cite[]{S20}. Among the different types of multiphase flows, a particularly interesting set corresponds to the so-called particle-laden suspensions in which small, immiscible and typically dilute particles are immersed in a carrier fluid (for instance, fine aerosol particles in air). The dynamics of gas-solid flows is rich and extraordinarily complex \cite[]{G94,J00,KH01,F12,TS14,FH17,LTSC20} so their understanding poses a great challenge. Even the study of granular flows in which the effect of interstitial fluid is neglected \cite[]{C90,G03,RN08,BP04,G19} entails enormous difficulties.

In the case that the particle-laden suspensions are dominated by collisions \cite[]{S20}, the extension of the classical kinetic theory of gases \cite[]{CC70,FK72,RL77} to gas-solid systems for relatively massive particles (high Stokes numbers) can be considered as an appropriate tool to model these systems. In this context and assuming nearly instantaneous collisions, the influence of gas-phase effects on the dynamics of solid particles is usually incorporated in the starting kinetic equation in an effective way via a fluid-solid interaction force \cite[]{K90,G94,J00}.
Some models for gas-solid suspensions \cite[]{LMJ91,TK95,SMTK96,WZL09,PS12,H13,WGZS14,SA17,ASG19,SA20} only consider the Stokes linear drag law for gas-solid interactions. Other models \cite[]{GTSH12} include also an additional Langevin-type stochastic term. In this case, based on the results obtained in direct numerical simulations (DNS), the impact of the viscous gas on solid particles in high-velocity---but low-Reynold numbers--- gas-solid flows is by means of a force constituted by three different terms: (i) a term proportional to the difference between the mean flow velocities of both phases, (ii) a drag force term proportional to the particle velocity, and (iii) a stochastic Langevin-like term taking into account the effects of neighbouring particles. While the second term mimics the dissipation of energy due to the friction of grains on the viscous gas, the third term models the energy gained by the solid particles due to their interaction with the particles of the interstitial gas. 

For small Knudsen numbers, the above suspension model \cite[]{GTSH12} has been solved by means of the Chapman--Enskog method \cite[]{CC70} adapted to dissipative dynamics. Explicit expressions for the Navier--Stokes--Fourier transport coefficients have been obtained in terms of the coefficient of restitution and the parameters of the suspension model \cite[]{GTSH12,GGG19a}. The knowledge of the forms of the transport coefficients has allowed to assess not only the impact of inelasticity on them [which was already analysed in the case of dry granular fluids \cite[]{BDKS98,GD99a}] but also the influence of the interstitial gas on the momentum and heat transport. Beyond the Navier--Stokes domain, this type of suspension models have been also considered to compute the rheological properties in sheared gas-solid suspensions [see for instance, \cite{TK95,SMTK96,PS12,H13,SMMD13,KIB14,ChVG15,SA17,HTG17,ASG19,HT19,SA20,GGG20,THSG20}].

The quantitative and qualitative accuracy of the (approximate) analytical results derived from the kinetic-theory two-fluid model \cite[]{GTSH12} have been confronted against computer simulations in several problems. In particular, the critical length for the onset of velocity vortices in the homogeneous cooling state of gas-solid flows obtained from a linear stability analysis presents an acceptable agreement with molecular dynamics (MD) simulations carried out for strong inelasticity \cite[]{GFHY16}. Simulations using a computational fluid dynamics (CFD) solver \cite[]{CD13,CDF15} of \cite{RS14} have shown a good agreement in the mean slip velocity with the kinetic-theory predictions \cite[]{FH16}. On the other hand, kinetic theory has been also assessed for describing clustering instabilities in sedimenting fluid-solid systems; good agreement is found at high solid-to-fluid  density ratios although the agreement is weaker for intermediate and low density ratios \cite[]{FLYH17}. In the case of non-Newtonian flows, the theoretical results \cite[]{SA17,ASG19,SA20} derived from the Stokes drag model for the ignited-quenched transition and the rheology of a sheared gas-solid suspension have been shown to compare very well with computer simulations. Regarding the Langevin-like model \cite[]{GTSH12}, the rheological properties of a moderately dense inertial suspension computed by a simpler version of this model exhibit a quantitative good agreement with MD simulations in the high-density region \cite[]{THSG20}. In addition, the extension to binary mixtures of this suspension model has been tested against Monte Carlo data and MD simulations for both time-dependent and steady homogeneous states with an excellent agreement \cite[]{KG14,GKG20,GGG21}.

In spite of the reliability of the generalised Langevin and Stokes drag models for capturing in an effective way the impact of gas phase on grains, it would be desirable to propose a suspension model that considers the real collisions between solid and gas particles. In the context of kinetic theory and as already mentioned in previous works \cite[]{GKG20}, a possibility would be to describe gas-solid flows in terms of a set of two coupled kinetic equations for the one-particle velocity distribution functions of the solid and gas phases. Nevertheless, the determination of the transport coefficients of the solid particles starting from the above suspension model is a very intricate problem. A possible way of overcoming the difficulties inherent to the description of gas-solid flows when one attempts to involve the different types of collisions is to assume that the properties of the gas phase are unaffected by the presence of solid particles. In fact, although sometimes not explicitly stated, this is one of the overarching assumptions in most of the suspension models reported in the granular literature. This assumption can be clearly justified in the case of particle-laden suspensions where the granular particles (or ``granular gas'') are sufficiently rarefied (dilute particles) and hence, the properties of the interstitial fluid can be supposed to be constant. This means that the background gas can be treated as a \emph{thermostat} at a constant temperature $T_g$.

Under these conditions and inspired in a paper of \cite{BMP02a}, we propose here the following suspension model. We consider a set of \emph{granular} particles immersed in a bath of elastic particles (\emph{molecular} gas) at equilibrium at a certain temperature $T_g$. While the collisions between granular particles are \emph{inelastic} (and characterised by a constant coefficient of normal restitution $\al$), the collisions between the granular and gas particles are considered to be elastic. In the homogeneous steady state, the energy lost by the solid particles due to their collisions among themselves is exactly compensated for by the energy gained by the grains due to their elastic collisions with particles of the molecular gas. In other words, the gas of \emph{inelastic} hard spheres (granular gas) is thermostated by a bath of \emph{elastic} hard spheres. The dynamic properties of this system in homogeneous steady states were studied years ago independently by \cite{BMP02a} and \cite{S03a}. Our goal here is going beyond the homogeneous state and determine the transport coefficients of the granular gas immersed in the molecular gas when the magnitude of the spatial gradients is small (Navier--Stokes domain).

It is quite apparent that this suspension model (granular particles plus molecular gas) can be seen as a binary mixture in which the concentration of one of the species (tracer species or granular particles) is much smaller than the other one (excess species or molecular gas). In these conditions, it is reasonable to assume that the state of the background gas (excess species) is not perturbed by the presence of the tracer species (granular particles). In addition, although the density of grains is very small, we will take into account not only the collisions between solid and gas particles, but also the grain-grain collisions in the kinetic equation of the one-particle distribution function $f(\mathbf{r},\mathbf{v};t)$ of solid particles. In spite of the simplicity of the model, it can be considered sufficiently robust since it retains most of the basic features of gas-solid flows such as the competition between the different spatial and time scales. As an added value and in contrast to the usual suspension models reported in the literature \cite[]{K90,G94,J00}, the model incorporates a new parameter: the ratio between the mass $m$ of the granular particles and the mass $m_g$ of the particles of the molecular gas.

As mentioned before, the main goal of the paper is to determine the Navier--Stokes--Fourier transport coefficients of the granular particles thermostated by a bath of elastic hard spheres. For a low-density granular gas, the distribution function $f(\mathbf{r},\mathbf{v};t)$ verifies the Boltzmann kinetic equation. More specifically, since granular particles collides among themselves and with particles of the molecular gas, the time evolution of the distribution $f$ involves the Boltzmann $J[f,f]$ and Boltzmann--Lorentz $J_g[f,f_g]$ collisions operators. Here, $f_g$ is the one-particle distribution function of the molecular gas. While the (nonlinear) collision operator $J$ accounts for the rate of change of $f$ due to inelastic collisions, the (linear) operator $J_g$ accounts for the rate of change of $f$ due to the elastic collisions between grains and gas particles. Interestingly, in the Brownian limiting case ($m\gg m_g$), the Boltzmann-Lorentz operator reduces to the Fokker--Planck operator so that, the results derived here reduce to those previously obtained by \cite{GGG19a} in this limiting case.

As in previous works \cite[]{GChV13a,GGG19a}, the transport coefficients are determined by solving the Boltzmann equation by the generalisation of the conventional Chapman--Enskog expansion \cite[]{CC70} to inelastic gases \cite[]{BP04,G19}. An important point in the perturbation method is the choice of the reference base state (zeroth-order approximation $f^{(0)}$). In the case of dry granular gases (no gas phase) and in the absence of spatial gradients, the solution $f^{(0)}(\mathbf{r},\mathbf{v}; t)$ to the Boltzmann equation is the \emph{local} version of the so-called homogeneous cooling state (HCS). On the other hand, in the case of granular suspensions, although one is interested in computing transport in steady conditions, the presence of the background molecular gas may induce a \emph{local} energy unbalance between the energy lost due to inelastic collisions and the energy transfer via elastic collisions. This leads in general to a non-stationary zeroth-order distribution $f^{(0)}$. Thus, as already did in previous calculations \cite[]{GChV13a,GGG19a}, we have to consider first the time-dependent distribution $f^{(0)}(\mathbf{r},\mathbf{v}; t)$ in order to arrive to the linear integral equations obeying the Navier--Stokes--Fourier transport coefficients. Then, to get explicit forms for the transport coefficients, the above integral equations are (approximately) solved under steady state conditions.

The plan of the paper is as follows. The Boltzmann kinetic equation for a granular gas immersed in a molecular gas is presented in section \ref{sec2} along with the corresponding balance equations for the densities of mass, momentum, and energy. The Brownian limit ($m/m_g\to \infty$) is also considered; it is shown that the Boltzmann--Lorentz operator $J_g[\mathbf{v}|f,f_g]$ reduces in this limiting case to the Fokker--Planck operator \cite[]{RL77,M89}, which is the basis of the Langevin-like suspension model \cite[]{GTSH12}. Section \ref{sec3} is devoted to the study of the homogeneous steady state (HSS). Although the HSS was already analysed by \cite{S03a} for a three-dimensional system ($d=3$), we revisit here this study by extending the analysis to an arbitrary number of dimensions $d$. As usual, the first-Sonine approximation to the velocity distribution $f(\mathbf{v})$ is considered to determine the temperature ratio $T/T_g$ and the fourth cumulant (or kurtosis) $a_2$ in terms of the parameter space of the system: the dimensionality $d$, the coefficient of restitution $\al$, the mass ratio $m/m_g$, the volume fraction $\phi$, and the (reduced) background temperature $T_g^*$. In the above Sonine solution, only linear terms in $a_2$ are considered. The theoretical results are compared against Monte Carlo simulations for $d=3$, $\phi=0.001$, $T_g^*=1000$, and different values of the mass ratio. The comparison shows in general an excellent agreement for the temperature ratio; some small quantitative discrepancies are found for $a_2$ in the case $m/m_g=1$.

Section \ref{sec4} addresses the application of the Chapman--Enskog-like expansion \cite[]{CC70} to the Boltzmann kinetic equation. Since the system is slightly disturbed from the HSS, the expansion is around the  \emph{local} version of the homogeneous state. However, as said before, for general small deviations from the homogeneous state the zeroth-order (reference) distribution function $f^{(0)}(\mathbf{r}, \mathbf{v};t)$ is a time-dependent distribution. As usual for elastic collisions \cite[]{CC70,FK72}, the Navier--Stokes--Fourier transport coefficients are in given in terms of the solutions of a set of coupled linear integral equations. On the other hand, due to the mathematical difficulties involved in the time-dependent problem, as in previous works \cite[]{GChV13a,GGG19a} the general results are restricted to steady-state conditions, namely, when the constraint $\zeta^{(0)}+\zeta_g^{(0)}=0$ holds at any point of the system. Here, $\zeta^{(0)}$ and $\zeta_g^{(0)}$ are the zeroth-order contributions to the partial production rates due to solid-solid and solid-gas collisions, respectively.
In the steady state, explicit expressions of the Navier--Stokes--Fourier transport coefficients are obtained in section \ref{sec5} by considering the leading terms in a Sonine polynomial expansion. As in the case of the temperature ratio and the kurtosis, in dimensionless form, the transport coefficients are provided in terms of $\al$, $m/m_g$, $\phi$, and $T_g^*$. An interesting result is that the expressions of the transport coefficients  reduce to those previously obtained by \cite{GGG19a} in the Brownian limit ($m/m_g\to \infty$). As an application of the results reported in section \ref{sec5}, a linear stability analysis of the HSS is carried out in section \ref{sec6}. Analogously to the analysis performed in the Brownian limit \cite[]{GGG19a}, the present analysis shows that the HSS is linearly stable regardless the value of the mass ratio $m/m_g$. The paper is ended in section \ref{sec7} with a brief summary of the results reported here.

\section{Boltzmann kinetic equation for a granular gas surrounded by a molecular gas}
\label{sec2}

\begin{figure}
\centering
\includegraphics[width=.35\columnwidth]{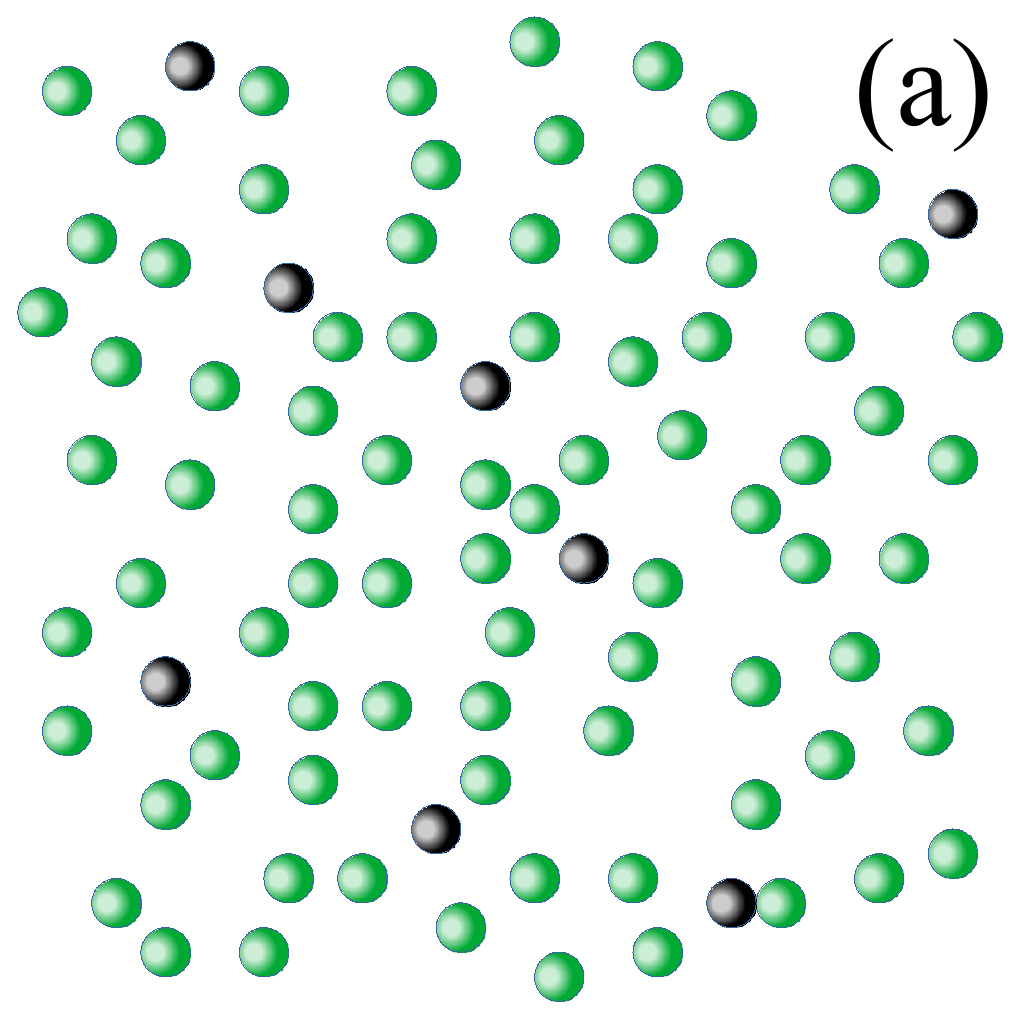}\hspace{5mm}\includegraphics[width=.35\columnwidth]{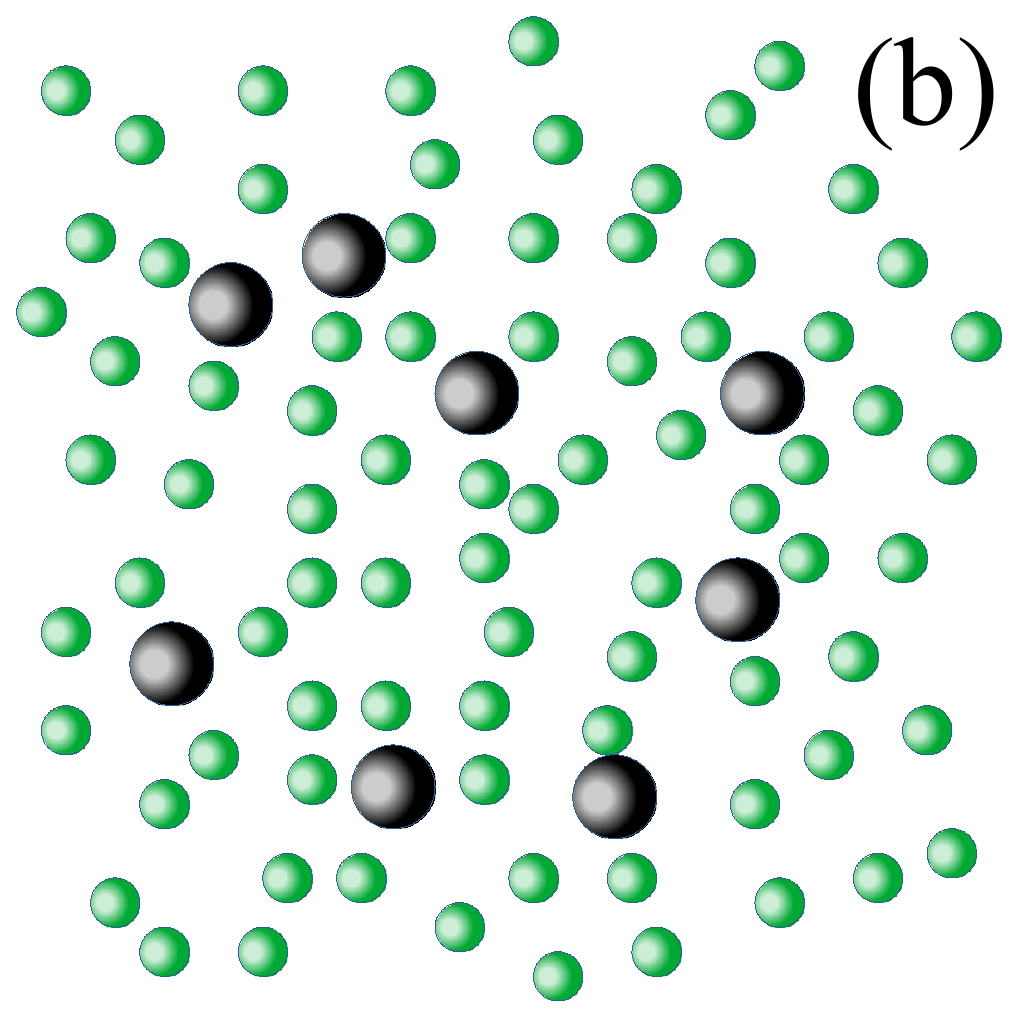}\vspace{5mm}\\
\includegraphics[width=.35\columnwidth]{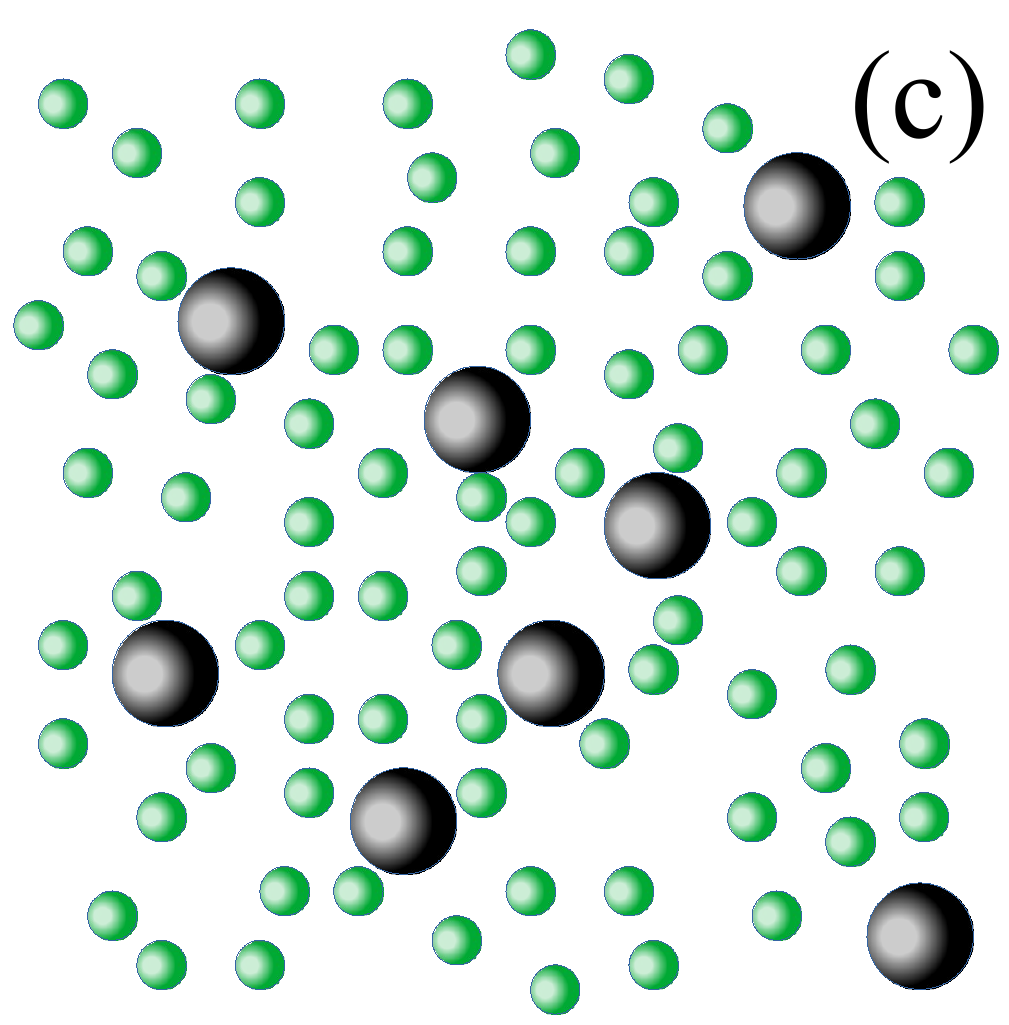}\hspace{5mm}\includegraphics[width=.35\columnwidth]{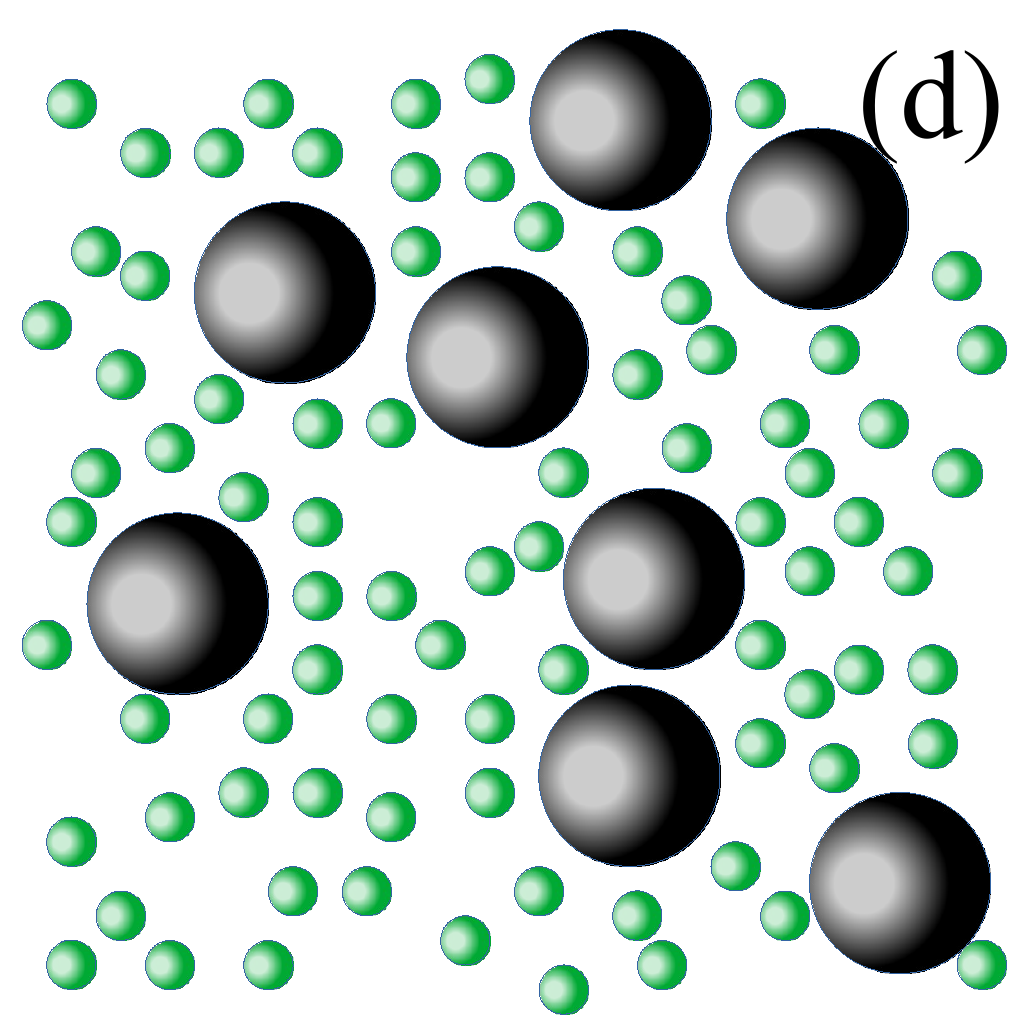}
\caption{Schematic plot of granular particles immersed in a molecular gas. Four different mass ratios are represented: 
 $m/m_g=1$ (a), $m/m_g=5$ (b), $m/m_g=10$ (c), and $m/m_g=50$ (d). Particles density $m/\sigma^3$ is constant in all the panels. Number density ratio is kept fixed to $n_g/n=10$.}
\label{figsc}
\end{figure}

We consider a gas of inelastic hard disks ($d=2$) or spheres ($d=3$) of mass $m$ and diameter $\sigma$. The spheres are assumed to be perfectly smooth so that, collisions between pairs are characterised by a (positive) constant coefficient of normal restitution $\al \leq 1$. When $\al=1$ ($\al<1$), the collisions are elastic (inelastic). The granular gas is immersed in a molecular gas constituted by hard disks or spheres of mass $m_g$ and diameter $\sigma_g$. Collisions between granular particles and gas molecules are considered to be \emph{elastic}. As discussed in section \ref{sec1}, we are interested here in describing a situation where the granular gas is sufficiently rarefied (the number density of granular particles is much smaller than that of the molecular gas) so that, the state of the molecular gas is not affected by the presence of solid (grains) particles. In this sense, the background (molecular) gas may be treated as a \emph{thermostat}, which is at equilibrium at the temperature $T_g$. Thus, the velocity distribution function $f_g$ of the molecular gas is the Maxwell--Boltzmann distribution:
\beq
\label{1.1}
f_g(\mathbf{V}_g)=n_g \Big(\frac{m_g}{2\pi T_g}\Big)^{d/2} \exp \Bigg(-\frac{m_g V_g^2}{2T_g}\Bigg),
\eeq
where $n_g$ is the number density of the molecular gas, $\mathbf{V}_g=\mathbf{v}-\mathbf{U}_g$, and $\mathbf{U}_g$ is the mean flow velocity of the molecular gas. Figure \ref{figsc} shows a schematic diagram of the system modelled in this work.

In the low-density regime, the time evolution of the one-particle velocity distribution function $f(\mathbf{r}, \mathbf{v}, t)$ of the granular gas is given by the Boltzmann kinetic equation. Since the granular particles collide among themselves and with the particles of the molecular gas, in the absence of external forces the velocity distribution $f$ verifies the kinetic equation
\beq
\label{1.2}
\frac{\partial f}{\partial t}+\mathbf{v}\cdot \nabla f=J[f,f]+J_g[f,f_g].
\eeq
Here, the Boltzmann collision operator $J[f,f]$ gives the rate of change of the distribution $f$ due to binary \emph{inelastic} collisions between granular particles. On the other hand, the Boltzmann-Lorentz operator $J_g[f,f_g]$ accounts for the rate of change of the distribution $f$ due to \emph{elastic} collisions between granular and molecular gas particles.

The explicit form of the nonlinear Boltzmann collision operator $J[f,f]$ is \cite[]{G19}
\beq
\label{1.3}
J[\mathbf{v}_1|f,f]=\sigma^{d-1}\int d\mathbf{v}_{2}\int d\widehat{\boldsymbol {\sigma
}}\Theta (\widehat{{\boldsymbol {\sigma }}} \cdot {\mathbf g}_{12}) (\widehat{\boldsymbol {\sigma }}\cdot {\mathbf g}_{12})
\left[\al^{-2}f({\mathbf v}_{1}'')f({\mathbf v}_{2}'')-f({\mathbf v}_{1})f({\mathbf v}_{2})\right],
\eeq
where $\mathbf{g}_{12}=\mathbf{v}_1-\mathbf{v}_2$ is the relative velocity, $\widehat{\boldsymbol {\sigma }}$ is a unit vector along the line of centers of the two spheres at contact, and $\Theta$ is the Heaviside step function. In Eq.\ \eqref{1.3}, the double primes denote pre-collisional velocities. The relationship between pre-collisional $(\mathbf{v}_1'', \mathbf{v}_2'')$ and post-collisional $(\mathbf{v}_1, \mathbf{v}_2)$ velocities is
\beq
\label{1.4}
\mathbf{v}_1''=\mathbf{v}_1-\frac{1+\al}{2\al}(\widehat{{\boldsymbol {\sigma }}} \cdot {\mathbf g}_{12})\widehat{\boldsymbol {\sigma }}, \quad
\mathbf{v}_2''=\mathbf{v}_2+\frac{1+\al}{2\al}(\widehat{{\boldsymbol {\sigma }}} \cdot {\mathbf g}_{12})\widehat{\boldsymbol {\sigma }}.
\eeq

The form of the linear Boltzmann-Lorentz collision operator $J_g[f,f_g]$ is \cite[]{G19,RL77}
\beq
\label{1.5}
J_g[\mathbf{v}_1|f,f_g]=\overline{\sigma}^{d-1}\int d\mathbf{v}_{2}\int d\widehat{\boldsymbol {\sigma
}}\Theta (\widehat{{\boldsymbol {\sigma }}} \cdot {\mathbf g}_{12}) (\widehat{\boldsymbol {\sigma }}\cdot {\mathbf g}_{12})
\left[f({\mathbf v}_{1}'')f_g({\mathbf v}_{2}'')-f({\mathbf v}_{1})f_g({\mathbf v}_{2})\right],
\eeq
where $\overline{\sigma}=(\sigma+\sigma_g)/2$. In Eq.\ \eqref{1.5}, the relationship between $(\mathbf{v}_1'',\mathbf{v}_2'')$ and $(\mathbf{v}_1,\mathbf{v}_2)$ is
\beq
\label{1.6}
\mathbf{v}_1''=\mathbf{v}_1-2 \mu_g(\widehat{{\boldsymbol {\sigma }}} \cdot {\mathbf g}_{12})\widehat{\boldsymbol {\sigma }}, \quad
\mathbf{v}_2''=\mathbf{v}_2+2 \mu (\widehat{{\boldsymbol {\sigma }}} \cdot {\mathbf g}_{12})\widehat{\boldsymbol {\sigma }},
\eeq
where $\mu_g=m_g/(m+m_g)$ and $\mu=m/(m+m_g)$. 

The relevant hydrodynamic fields of the granular gas are the number density $n(\mathbf{r}; t)$, the mean flow velocity $\mathbf{U}(\mathbf{r}; t)$, and the granular temperature $T(\mathbf{r}; t)$. They are defined, respectively, as
\beq
\label{1.7}
\left\{n, n \mathbf{U}, n T\right\}=\int d\mathbf{v} \left\{1, \mathbf{v}, m V^2\right\} f(\mathbf{v}),
\eeq
where $\mathbf{V}=\mathbf{v}-\mathbf{U}$ is the peculiar velocity. In general, the mean flow velocity $\mathbf{U}$ of solid particles is different from the mean flow velocity $\mathbf{U}_g$ of molecular gas particles. As we will show later, the difference $\mathbf{U}-\mathbf{U}_g$ induces a nonvanishing contribution to the heat flux.

The macroscopic balance equations for the granular gas are obtained by multiplying Eq.\ \eqref{1.2} by $\left\{1, \mathbf{v}, m V^2\right\}$ and integrating over velocity. The result is
\beq
\label{1.8}
D_t n+n\nabla\cdot \mathbf{U}=0,
\eeq
\beq
\label{1.9}
\rho D_t\mathbf{U}=-\nabla \cdot \mathsf{P}+\boldsymbol{\mathcal{F}}[f],
\eeq
\beq
\label{1.10}
D_tT+\frac{2}{dn}\Big(\nabla \cdot \mathbf{q}+\mathsf{P}:\nabla \mathbf{U}\Big)=- T \zeta-T \zeta_g.
\eeq
Here, $D_t=\partial_t+\mathbf{U}\cdot \nabla$ is the material derivative, $\rho=m n$ is the mass density of solid particles, and the pressure tensor $\mathsf{P}$ and the heat flux vector $\mathbf{q}$ are given, respectively, as
\beq
\label{1.11}
\mathsf{P}=\int d\mathbf{v}\; m \mathbf{V}\mathbf{V} f(\mathbf{v}),
\eeq
\beq
\label{1.11.1}
\mathbf{q}=\int d\mathbf{v}\; \frac{m}{2} V^2 \mathbf{V} f(\mathbf{v}).
\eeq
Since the Boltzmann--Lorentz collision term $J_g[f,f_g]$ does not conserve momentum, then the production of momentum $\boldsymbol{\mathcal{F}}[f]$ is in general different from zero. It is defined as
\beq
\label{1.12}
\boldsymbol{\mathcal{F}}[f]=\int d\mathbf{v}\; m \mathbf{V}  J_g[f,f_g]=-\frac{2\pi^{(d-1)/2}}{\Gamma\Big(\frac{d+3}{2}\Big)} m_g \mu \overline{\sigma}^{d-1}\int d\mathbf{V}_1\int d\mathbf{V}_2\; g_{12}\mathbf{g}_{12}\; f(\mathbf{V}_1) f_g(\mathbf{V}_2).
\eeq
In addition, the partial production rates $\zeta$ and $\zeta_g$ are given, respectively, as
\beq
\label{1.13}
\zeta=-\frac{m}{d n T}\int d\mathbf{v}\; V^2\; J[\mathbf{v}|f,f],
\eeq
\beq
\label{1.14}
\zeta_g=-\frac{m}{d n T}\int d\mathbf{v}\; V^2\; J_g[\mathbf{v}|f,f_g].
\eeq
The cooling rate $\zeta$ gives the rate of kinetic energy loss due to inelastic collisions between particles of the granular gas. It vanishes for inelastic collisions. The term $\zeta_g$ gives the transfer of kinetic energy between the particles of the granular and molecular gas.
It vanishes when the granular and molecular gas are at the same temperature ($T_g=T$).

\subsection{Brownian limit ($m/m_g \to \infty$)}

The suspension model defined by the Boltzmann equation \eqref{1.2} applies in principle for arbitrary values of the mass ratio $m/m_g$. On the other hand, a physically interesting situation arises in the so-called Brownian limit, namely, when the granular particles are much heavier than the particles of the surrounding molecular gas ($m/m_g\to \infty$). In this case, a Kramers--Moyal expansion \cite[]{RL77,RSD83,M89} in the velocity jumps $\delta \mathbf{v}=(2/(1+m/m_g)) (\widehat{\boldsymbol {\sigma }}\cdot {\mathbf g}_{12}) \mathbf{g}_{12}$ allows us to approximate the Boltzmann--Lorentz operator $J_g[\mathbf{v}|f,f_g]$ by the Fokker--Planck operator $J_g^{\text{FP}}[\mathbf{v}|f,f_g]$ \cite[]{RL77,RSD83,M89,BDS99,SVCP10}:
\beq
\label{1.15}
J_g[f,f_g]\to J_g^{\text{FP}}[f,f_g]=\gamma \frac{\partial}{\partial \mathbf{v}}\cdot \Bigg(\mathbf{v}+\frac{T_g}{m}\frac{\partial}{\partial \mathbf{v}}\Bigg)f(\mathbf{v}),
\eeq
where the friction coefficient $\gamma$ is defined as
\beq
\label{1.16}
\gamma=\frac{4\pi^{(d-1)/2}}{d\Gamma\Big(\frac{d}{2}\Big)}\Big(\frac{m_g}{m}\Big)^{1/2}\Bigg(\frac{2T_g}{m}\Bigg)^{1/2}n_g \overline{\sigma}^{d-1}.
\eeq
Upon obtaining Eqs.\ \eqref{1.15}--\eqref{1.16}, it has been assumed that $\mathbf{U}_g=\mathbf{0}$ and a Maxwellian distribution for the distribution $f(\mathbf{v})$ of the granular gas.

Most of the suspension models employed in the granular literature to fully account for the influence of
the interstitial molecular fluid on the dynamics of grains are based on the replacement of $J_g[f,f_g]$ by the Fokker--Planck operator \eqref{1.15} \cite[]{KH01}. More specifically, for general inhomogeneous states, the impact of the background molecular gas on solid particles
is through an effective force composed by three different terms: (i) a term proportional to the difference $\Delta \mathbf{U}=\mathbf{U}-\mathbf{U}_g$, (ii) a drag force term mimicking the friction of grains on the viscous interstitial gas, and (iii) a stochastic Langevin-like term accounting for the energy gained by grains due to their interactions with particles of the molecular gas (neighbouring particles effect) \cite[]{GTSH12}. This yields the following kinetic equation for gas-solid suspensions \footnote{There are three different scalars $(\beta,\gamma,\xi)$ in the suspension model proposed by \cite{GTSH12}; each one of the coefficients is associated with the different terms of the fluid-solid force. For the sake of simplicity, the results derived by \cite{GGG19a} were obtained by assuming that $\beta=\gamma=\xi$.}
\beq
\label{1.17}
\frac{\partial f}{\partial t}+\mathbf{v}\cdot \nabla f-\gamma\Delta \mathbf{U}\cdot \frac{\partial f}{\partial \mathbf{v}}-\gamma\frac{\partial}{\partial \mathbf{v}}\cdot \mathbf{V} f-\gamma \frac{T_{g}}{m}\frac{\partial^2 f}{\partial v^2}=J[\mathbf{v}|f,f].
\eeq
The Boltzmann equation \eqref{1.17} has been solved by means of the Chapman--Enskog method \cite[]{CC70} to first-order in spatial gradients. Explicit forms for the Navier--Stokes--Fourier transport coefficients have been obtained in steady-state conditions, namely, when the cooling terms are compensated for by the energy gained by the solid particles due to their collisions with the bath particles \cite[]{GChV13a,GGG19a}. Thus, the results derived in the present paper must be consistent with those previously obtained by \cite{GGG19a} when the limit $m_g/m\to 0$ is considered in our general results.

\section{Homogeneous steady state}
\label{sec3}

As a first step and before studying inhomogeneous states, we consider the HSS. The HSS is the reference base state (zeroth-order approximation) used in the Chapman--Enskog perturbation method \cite[]{CC70}. Therefore, its investigation is of great importance. The HSS was widely analysed by \cite{S03a} for a three-dimensional granular gas. Here, we extend these calculations to a general dimension $d$.

In the HSS, the density $n$ and temperature $T$ are spatially uniform, and with an appropriate selection of the frame reference, the mean flow velocities vanish ($\mathbf{U}=\mathbf{U}_g=\mathbf{0}$). Consequently, the Boltzmann equation \eqref{1.2} reads
\beq
\label{2.1}
\frac{\partial f}{\partial t}=J[f,f]+J_g[f,f_g].
\eeq
Moreover, the velocity distribution $f(\mathbf{v})$ of the granular gas is isotropic in $\mathbf{v}$ so that the production of momentum $\boldsymbol{\mathcal{F}}[f]=\mathbf{0}$, according to Eq.\ \eqref{1.12}. Thus, the only nontrivial balance equation is that of the temperature \eqref{1.10}:
\beq
\label{2.2}
\frac{\partial \ln T}{\partial t}=-\left(\zeta+\zeta_g\right).
\eeq
As mentioned in section \ref{sec2}, since collisions between granular particles are inelastic, then the cooling rate $\zeta>0$. Collisions between particles of granular and molecular gases are elastic and so, the total kinetic energy of two colliding particles is conserved. On the other hand, since in the steady state the background gas acts as a thermostat, then the mean kinetic energy of granular particles is smaller than that of the molecular gas and so, $T<T_g$. This necessarily implies that $\zeta_g<0$. Therefore, in the \emph{steady} state, the terms $\zeta$ and $\zeta_g$ exactly compensates each other and one gets the steady-state condition
\beq
\label{2.3}
\zeta+\zeta_g=0.
\eeq
The condition \eqref{2.3} allows one to get the steady granular temperature $T$. However, according to the definitions \eqref{1.13} and \eqref{1.14}, the determination of $\zeta$ and $\zeta_g$ requires to know the velocity distribution $f(\mathbf{v})$. For inelastic collisions ($\al\neq 1$), to date the solution of the Boltzmann equation \eqref{2.1} has not been found. On the other hand, a good estimate of $\zeta$ and $\zeta_g$ can be obtained when the first-Sonine approximation to $f$ is considered \cite[]{BP04}. In this approximation, $f(\mathbf{v})$ is given by
\beq
\label{2.4}
f(\mathbf{v})\simeq f_{\text{MB}}(\mathbf{v})\Bigg\{1+\frac{a_2}{2} \Bigg[\Bigg(\frac{m v^2}{2T}\Bigg)^2-(d+2)\frac{mv^2}{2T}+\frac{d(d+2)}{4}\Bigg]\Bigg\},
\eeq
where
\beq
\label{2.5}
f_{\text{MB}}(\mathbf{v})=n \Big(\frac{m}{2\pi T}\Big)^{d/2} \exp \Big(-\frac{m v^2}{2T}\Big)
\eeq
is the Maxwell--Boltzmann distribution and
\beq
\label{2.6}
a_2=\frac{1}{d(d+2)}\frac{m^2}{n T^2}\int d\mathbf{v}\; v^4\; f(\mathbf{v})-1
\eeq
is the kurtosis or fourth cumulant. This quantity measures the departure of the distribution $f(\mathbf{v})$ from its Maxwellian form $f_{\text{MB}}(\mathbf{v})$. From experience with the dry granular case \cite[]{NE98,GD99a,MS00,SM09}, the magnitude of the cumulant $a_2$ is expected to be very small and so, the Sonine approximation \eqref{2.4} to the distribution $f$ turns out to be reliable. In the case that $|a_2|$ does not remain small for high inelasticity, one should include cumulants of higher order in the Sonine polynomial expansion of $f$. However, the possible lack of convergence of the Sonine polynomial expansion for very small values of the coefficient of restitution \cite[]{BP06a,BP06b} puts on doubt the reliability of the Sonine expansion in the high inelasticity region. Here, we will restrict to values of $\al$ where $|a_2|$ remains relatively small. 

The expressions of $\zeta$ and $\zeta_g$ can be now obtained by replacing in Eqs.\ \eqref{1.13} and \eqref{1.14} $f$ by its Sonine approximation \eqref{2.4}. Retaining only linear terms in $a_2$, the forms of the dimensionless production rates
\beq
\label{2.7}
\zeta^*=\frac{\ell \zeta}{v_\text{th}}, \quad  \zeta_g^*=\frac{\ell \zeta_g}{v_\text{th}}
\eeq
can be written as \cite[]{NE98,BP06a}
\beq
\label{2.8}
\zeta^*=\widetilde{\zeta}^{(0)}+\widetilde{\zeta}^{(1)} a_2, \quad \zeta_g^*=\widetilde{\zeta}_g^{(0)}+\widetilde{\zeta}_g^{(1)} a_2,
\eeq
where
\beq
\label{2.9}
\widetilde{\zeta}^{(0)}=\frac{\sqrt{2}\pi^{(d-1)/2}}{d\Gamma\Big(\frac{d}{2}\Big)}(1-\al^2), \quad \widetilde{\zeta}^{(1)}=\frac{3}{16} \widetilde{\zeta}^{(0)},
\eeq
\beq
\label{2.10}
\widetilde{\zeta}_g^{(0)}=2x(1-x^2) \Big(\frac{\mu T}{T_g}\Big)^{1/2}\gamma^*, \quad \widetilde{\zeta}_g^{(1)}=\frac{\mu_g}{8}x^{-3}\Big[x^2\left(4-3\mu_g\right)-\mu_g\Big]
\Big(\frac{\mu T}{T_g}\Big)^{1/2}\gamma^*.
\eeq
Here, $\ell=1/(n\sigma^{d-1})$ is proportional to the mean free path of hard spheres, $v_\text{th}=\sqrt{2T/m}$ is the thermal velocity, and
we have introduced the auxiliary parameters
\beq
\label{2.11}
x=\Bigg(\mu_g+\mu \frac{T_g}{T}\Bigg)^{1/2},
\eeq
and
\beq
\label{2.12}
\gamma^*=\varepsilon \; \Bigg(\frac{T_g}{T}\Bigg)^{1/2}, \quad \varepsilon=\frac{\ell \gamma}{\sqrt{2T_g/m}}=\frac{\sqrt{2}\pi^{d/2}}{2^d d \Gamma\left(\frac{d}{2}\right)}\frac{1}{\phi \sqrt{T_g^*}}.
\eeq
Here,
\beq
\label{2.13}
\phi=\frac{\pi^{d/2}}{2^{d-1}d\Gamma \left(\frac{d}{2}\right)}n\sigma^d
\eeq
is the solid volume fraction and
\beq
\label{2.14}
T_g^*=\frac{T_g}{m\sigma^2 \gamma^2}
\eeq
is the (reduced) bath temperature. The (reduced) friction coefficient $\gamma^*$ characterises the rate at which the collisions between grains and molecular particles occur. Equations \eqref{2.9} and \eqref{2.10} agree with those obtained by \cite{S03a} for $d=3$.

To close the problem, we have to determine the kurtosis $a_2$. In this case, one has to compute the collisional moments
\beq
\label{2.15}
\Lambda\equiv \int d\mathbf{v}\; v^4\; J[f,f], \quad \Lambda_g\equiv \int d\mathbf{v}\; v^4\; J_g[f,f_g].
\eeq
In the steady state, apart from Eq.\ \eqref{2.3}, one has the additional condition
\beq
\label{2.15.1}
\Lambda+\Lambda_g=0.
\eeq
The moments $\Lambda$ and $\Lambda_g$ have been obtained in previous works \cite[]{NE98,BP06a,GVM09,G19} by replacing $f$ by its first Sonine form \eqref{2.4} and neglecting nonlinear terms in $a_2$. In terms of the (reduced) friction coefficient $\gamma^*$, the expressions of
\beq
\label{2.16}
\left\{\Lambda^*,\Lambda_g^*\right\}=\frac{\ell}{n v_\text{th}^5}\left\{\Lambda,\Lambda_g\right\}
\eeq
are given by
\beq
\label{2.17}
\Lambda^*=\Lambda^{(0)}+\Lambda^{(1)} a_2, \quad \Lambda_g^*=\Lambda_g^{(0)}+\Lambda_g^{(1)} a_2,
\eeq
where
\beq
\label{2.18}
\Lambda^{(0)}=-\frac{\pi^{(d-1)/2}}{\sqrt{2}\Gamma\Big(\frac{d}{2}\Big)}\Big(d+\frac{3}{2}+\al^2\Big)(1-\al^2),
\eeq
\beq
\label{2.18.1}
\Lambda^{(1)}=-\frac{\pi^{(d-1)/2}}{\sqrt{2}\Gamma\Big(\frac{d}{2}\Big)}\Big[\frac{3}{32}\left(10d+39+10\al^2\right)+\frac{d-1}{1-\al}\Big](1-\al^2),
\eeq
\beq
\label{2.19}
\Lambda_g^{(0)}=d x^{-1}\left(x^2-1\right)\left[8 \mu_g x^4+x^2\left(d+2-8\mu_g\right)+\mu_g\right]\Bigg(\frac{\mu T}{T_g}\Bigg)^{1/2}\gamma^*,
\eeq
\beqa
\label{2.20}
\Lambda_g^{(1)}&=&\frac{d}{8} x^{-5}\Bigg\{4x^6\Big[30\mu_g^3-48\mu_g^2+3(d+8)\mu_g-2(d+2)\Big]+\mu_g x^4 \Big[-48\mu_g^2\nonumber\\
& & +3(d+26)\mu_g-8(d+5)\Big]+\mu_g^2 x^2\left(d+14-9\mu_g\right)-3\mu_g^3\Bigg\}\Bigg(\frac{\mu T}{T_g}\Bigg)^{1/2}\gamma^*.\nonumber\\
\eeqa
For hard spheres ($d=3$), Eqs.\ \eqref{2.18}--\eqref{2.20} are consistent with those previously obtained by \cite{S03a}.

Inserting Eqs.\ \eqref{2.8}--\eqref{2.10} and \eqref{2.17}--\eqref{2.20} into Eqs.\ \eqref{2.3} and \eqref{2.15.1}, respectively, one gets the set of coupled equations:
\beq
\label{2.21}
\widetilde{\zeta}^{(0)}+\widetilde{\zeta}_g^{(0)}+\Big(\widetilde{\zeta}^{(1)}+\widetilde{\zeta}_g^{(1)}\Big)a_2=0,
\eeq
\beq
\label{2.21.1}
\Lambda^{(0)}+\Lambda_g^{(0)}+\Big(\Lambda^{(1)}+\Lambda_g^{(1)}\Big)a_2=0.
\eeq
Eliminating $a_2$ in Eqs.\ \eqref{2.21} and \eqref{2.21.1}, one achieves the following closed equation for the temperature ratio $T/T_g$:
\beq
\label{2.22}
\Big(\widetilde{\zeta}^{(1)}+\widetilde{\zeta}_g^{(1)}\Big)\Big(\Lambda^{(0)}+\Lambda_g^{(0)}\Big)=
\Big(\widetilde{\zeta}^{(0)}+\widetilde{\zeta}_g^{(0)}\Big)\Big(\Lambda^{(1)}+\Lambda_g^{(1)}\Big).
\eeq
For given values of $\al$, $\phi$, and $T_g^*$, the numerical solution of Eq.\ \eqref{2.22} gives $T/T_g$. Once the temperature ratio is determined,
the cumulant $a_2$ is simply given by
\beq
\label{2.23}
a_2=-\frac{\widetilde{\zeta}^{(0)}+\widetilde{\zeta}_g^{(0)}}{\widetilde{\zeta}^{(1)}+\widetilde{\zeta}_g^{(1)}}=
-\frac{\Lambda^{(0)}+\Lambda_g^{(0)}}{\Lambda^{(1)}+\Lambda_g^{(1)}}.
\eeq

\subsection{Brownian limit}

Before illustrating the dependence of $T/T_g$ and $a_2$ on $\al$ for given values of $\phi$ and $T_g^*$, it is interesting to consider the Brownian limit $m_g/m\to 0$. In this limiting case, $\mu_g\to 0$, $\mu\to 1$, $x\to \sqrt{T_g/T}$, and so
\beq
\label{2.24}
\widetilde{\zeta}_g^{(0)}\to 2\Bigg(1-\frac{T_g}{T}\Bigg)\gamma^*, \quad \widetilde{\zeta}_g^{(1)}\to 0,
\eeq
\beq
\label{2.24.1}
\Lambda_g^{(0)}\to d(d+2)\Bigg(\frac{T_g}{T}-1\Bigg)\gamma^*, \quad
\Lambda_g^{(1)}\to -d(d+2)\gamma^*.
\eeq
Taking into these results, the set of equations \eqref{2.21} and \eqref{2.21.1} can be written in the Brownian limit as
\beq
\label{2.25}
2\gamma^*\Bigg(\frac{T_g}{T}-1\Bigg)=\zeta^*, \quad d(d+2)\left(\gamma^* a_2-\frac{1}{2}\zeta^*\right)=\Lambda^*.
\eeq
These equations are the same as those derived by \cite{GGG19a} [see Eqs.\ (29) and (34) of this paper] by using the suspension model \eqref{1.17}. This shows the consistency of the present results in the HSS with those obtained in the Brownian limit.


\subsection{DSMC simulations}
\label{DSMC}

The previous analytical results have been obtained by using the first-Sonine approximation \eqref{2.4} to $f$. Thus, it is worth solving the Boltzmann kinetic equation by means of an alternative method to test the reliability of the theoretical predictions for $T/T_g$ [Eq.\ \eqref{2.22}] and $a_2$ [Eq.\ \eqref{2.23}]. The Direct Simulation Monte Carlo (DSMC) method developed by \cite{B94} is considered here to numerically solve the Boltzmann equation in the homogeneous state. As described in section 1, we treat in this paper the molecular gas as a thermostat in the sense that its state is not perturbed by the presence of grains. Therefore, the collision stage in the DSMC method must be slightly modified to accurately reproduce Eq.\ \eqref{2.1}. We follow similar steps as proposed by \cite{mg02} to numerically solve the Boltzmann--Enskog equation of a homogeneous granular mixture.

The simulation is initiated by drawing the particle velocities from a Maxwellian distribution at temperature $T_g$ following the Box--Muller transform \cite[]{BM58}. Since the granular gas is assumed to be spatially homogeneous, only the collision stage is described here.  The procedure can be summarised as follows:
\begin{enumerate}
    \item A required number of $N_{j}^{\delta t}$ candidate pairs to collide in a time $\delta t$ is selected. This number is given by\footnote{In contrast to the work of \cite{mg02}, we consider here a very dilute system and so, the pair correlation functions are set equal to 1.} \cite[]{mg02}
    \beq\label{MC1}
    N_{j}^{\delta t}=j\frac{2^{d-3}d\Gamma\left(\frac{d}{2}\right)}{\pi^{d/2}}\frac{(\sigma+\sigma_{j})^2}{\sigma^d}N_j\phi g_{j}^\text{max}\delta t,
    \eeq
    where $j=1\ (j=2)$ refers to a granular (molecular) particle. Namely, $N_{1}^{\delta t}$ refers to granular-granular collisions, while $N_{2}^{\delta t}$ refers to granular-molecular collisions. Here, $N_j$ is the total number of particles of species $j$ and $g_{j}^\text{max}$ is an upper bound of the average relative velocity. A good estimate is $g_{j}^\text{max}=Cv^\text{th}_{j}$, where $v^\text{th}_{j}=\sqrt{2T_g/\overline{m}}$ is the mean thermal velocity, $\overline{m}=(m+m_j)/2$, and $C$ is a constant, e.g., $C = 5$ \cite[]{B94}. Note that in Eq. \eqref{MC1} collisions among molecular particles themselves have been neglected.
    
    \item A colliding direction $\widehat{{\boldsymbol {\sigma }}}_j$ is randomly selected with equiprobability.
    
    \item The collision is accepted if
    \beq
    \label{MC2}
    |\widehat{{\boldsymbol {\sigma }}}_j\cdot \mathbf{g}_{12}|= |\widehat{{\boldsymbol {\sigma }}}_j\cdot (\mathbf{v}_{1}-\mathbf{v}_2)|>U(0,1)g_{j}^\text{max},
    \eeq
    where $U(0,1)$ is a random number uniformly distributed in $[0,1]$.
    
    \item If the collision is accepted, only granular particles velocities are updated according to the relationships \eqref{1.4} for $j=1$ and \eqref{1.6} for $j=2$.
\end{enumerate}

The former procedure constitutes an intermediate method between Bird's \cite[]{B94} and Nanbu's \cite[]{N86} schemes since in the latter only one of the colliding particles changes its velocity. However, as pointed out by \cite{MS97}, both schemes are equivalent and equally useful to solve the Boltzmann equation since in both techniques momentum is conserved on average. Thus, we do not need to account for collisions among molecular particles themselves because $n/n_g\ll 1$ and the computational cost would be very expensive.

Moreover, in the theory all the mechanical information of the molecular gas (with the exception the mass ratio $m/m_g$) is enclosed in the (reduced) friction coefficient $\gamma^*$ throughout the reduced (bath) temperature $T^*_g$. Let us denote by $N_g$ and $N$  the total number of granular and molecular particles, respectively. Since $N/N_g=n/n_g$, then $\sigma$ and $\sigma_g$ are related by
\beq
\label{MC3}
\sigma_g=\left[\left(\frac{\sqrt{\pi}}{4\sqrt{2}}\frac{N}{N_g}\sqrt{\frac{m}{m_g}}\frac{1}{\phi\sqrt{T_g^*}}\right)^{1/(d-1)}-1\right]\sigma.
\eeq
Upon deriving Eq.\ \eqref{MC3} use has been made of the relationships 
\beq
\label{MC3.1}
n=\frac{2^{d-1}d\Gamma\left(\frac{d}{2}\right)}{\pi^{d/2}}\sigma^{-d}\phi, \quad 
n_g=\frac{d \Gamma\left(\frac{d}{2}\right)}{4\pi^{(d-1)/2}}\left(\frac{m}{m_g}\right)^{1/2}\left(\frac{m}{2 T_g}\right)^{1/2}\overline{\sigma}^{1-d}\gamma.
\eeq
Equation \eqref{MC3} establishes a constraint in the inputs regarding the molecular gas. For this reason, once the inputs appearing in the theory ($d,m/m_g,T^*_g,\phi$) are fixed, then we choose $N_g$ in such a way that $N_g/N\gg 1$ and $\sigma_g/\sigma> 0$.

\begin{figure}
\begin{center}
\includegraphics[width=.55\columnwidth]{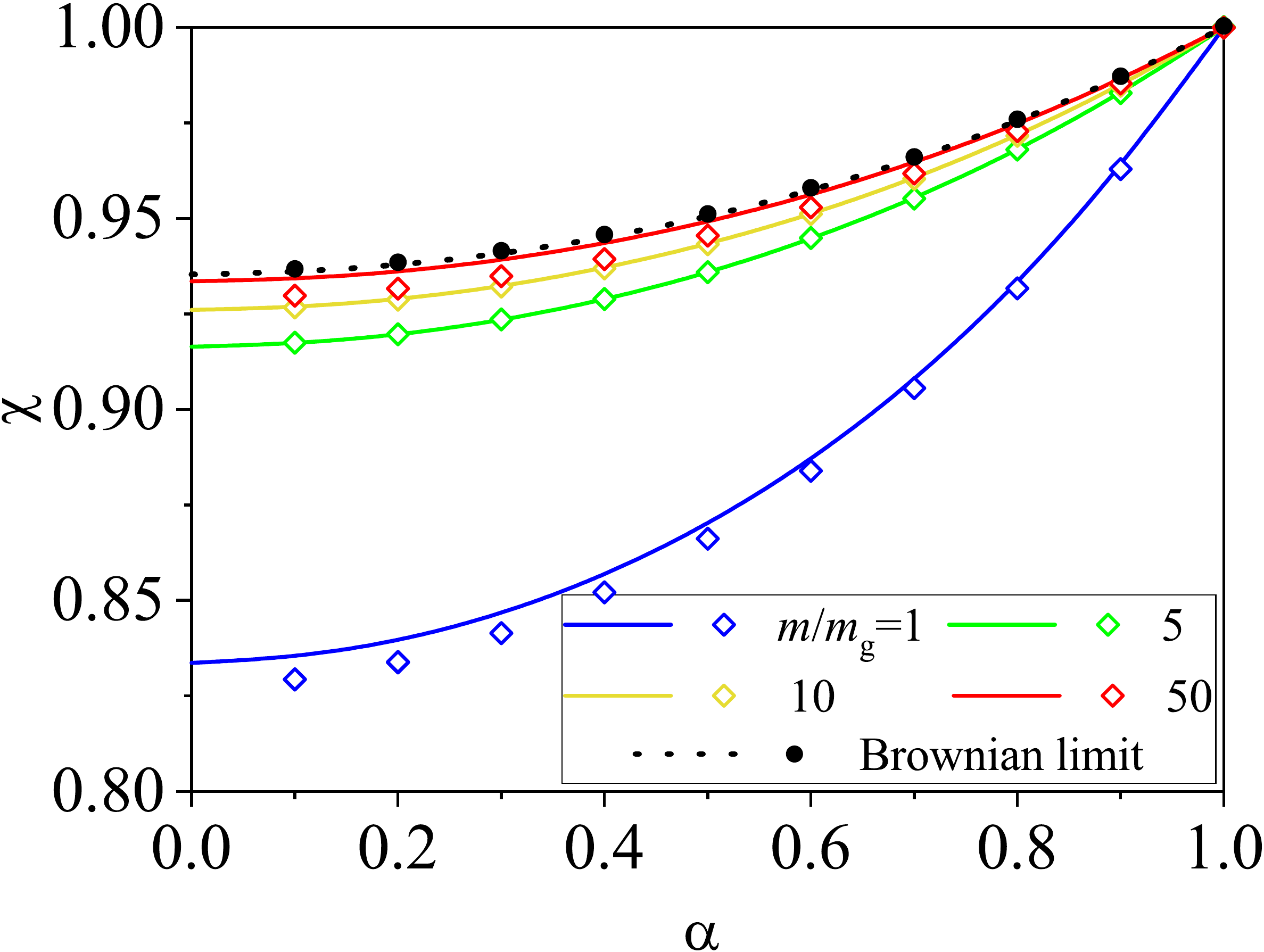}
\end{center}
\caption{Temperature ratio $\chi\equiv T/T_g$ versus the coefficient of normal restitution $\al$ for $d=3$, $\phi=0.001$, $T_g^*=1000$, and four different values of the mass ratio $m/m_g$ [from top to bottom, $m/m_g=50, 10, 5$, and 1]. The solid lines are the theoretical results obtained by numerically solving Eq.\ \eqref{2.22} and the symbols are the Monte Carlo simulation results. The dotted line is the result obtained by \cite{GGG19a} by using the Langevin-like suspension model \eqref{1.17} while black circles refer to DSMC simulations implemented using the time-driven approach \eqref{MC4}.
\label{fig1}}
\end{figure}

Figure \ref{fig1} shows the dependence of the temperature ratio $\chi\equiv T/T_g$ on the coefficient of restitution $\al$ for several values of the mass ratio $m/m_g$. A very dilute ensemble $(\phi=0.001)$ of hard spheres $(d=3)$ is considered. The value of the reduced (bath) temperature $T^*_g$ is selected so that collisions between grains and with the interstitial gas are both relevant on the dynamics of grains. Here, we chose $T_g^*=1000$. The lines are the theoretical results obtained by numerically solving Eq.\ \eqref{2.22}. The symbols refer to DSMC simulations performed following the method described above. Four different values of the mass ratio $m/m_g$ are considered ($m/m_g=1,5,10$, and 50). For the sake of comparison, the dotted line shows the $\al$-dependence of $\chi$ achieved in the Brownian limit, namely when considering the Langevin-like suspension model described in Eq. \eqref{1.17} \cite[]{GGG19a}. In addition, black circles refer to DSMC simulations performed by employing the model \eqref{1.17}. To carry on these kind of simulations, the influence of the external fluid on grains is taken into account by updating the velocity of every single grain each time step $\delta t$ according to \cite[]{KG14,GKG21}:
\beq
\label{MC4}
\mathbf{v}\to e^{-\gamma\delta t}\mathbf{v}+\left(\frac{6\gamma T_g \delta t}{m}\right)^{1/2}\mathbf{U}[-1,1].
\eeq
Here, $\mathbf{U}$ is an uniformly distributed random vector in $[-1,1]^3$. Equation \eqref{MC4} converges to the Fokker--Plank operator \eqref{1.15} when a time step $\delta t$ much smaller than the mean free time between collisions is considered \cite[]{KG14}.

Figure \ref{fig1} ensures the reliability of the results derived in this section for two different reasons: (i) a good agreement between theory and simulation is found and (ii) the convergence towards the Brownian limit can be clearly observed. Surprisingly, this convergence is fully reached for relatively small values of the mass ratio ($m/m_g\approx 50$) in contrast to the results reported by \cite{S03a}. Another unexpected result concerns the lack of energy equipartition ($T\ne T_g$) \cite[]{BT02,DHGD02}. One expects the temperature of the granular and molecular gases to be similar when the particles that composed them are mechanically comparable. However, according to Eq.\ \eqref{1.6}, the transmission of energy per individual collision from a molecular particle to a grain is bigger when their masses are similar. Nonetheless, the constraint imposed by Eq.\ \eqref{MC3} leads to a dependence of $N/N_g$ on the mass ratio $m/m_g$ for fixed $\sigma_g$. Thus, $N_g/N\propto m/m_g$ and so, the number density of the molecular gas increases as increasing the mass ratio. This way, the mean force exerted by the molecular particles on the grains is greater and therefore, the thermalisation caused by the presence of the interstitial fluid is much more effective. The steady temperature ratio $\chi$ is reached when the energy lost by collisions is compensated for by the energy provided by the bath. Hence, the nonequipartition of energy turns out then to be remarkable to small values of $m/m_g$ and $\alpha$. The former contrasts again with the results plotted in Fig.\ 2. of \cite{S03a}. However, as discussed in this section, for hard spheres ($d=3$) the results obtained in the HSS are consistent with those reported by \cite{S03a}. Consequently, the discrepancies found both in the convergence to the Brownian limit and in the dependence of the energy nonequipartition on the mass ratio are just a matter of the way of scaling the variables. In our study, we have introduced $\gamma^*$ in Eq.\ \eqref{2.21} as an auxiliary dimensionless variable for the sake of comparison with the results obtained by \cite{GGG19a}.

\begin{figure}
\begin{center}
\includegraphics[width=.55\columnwidth]{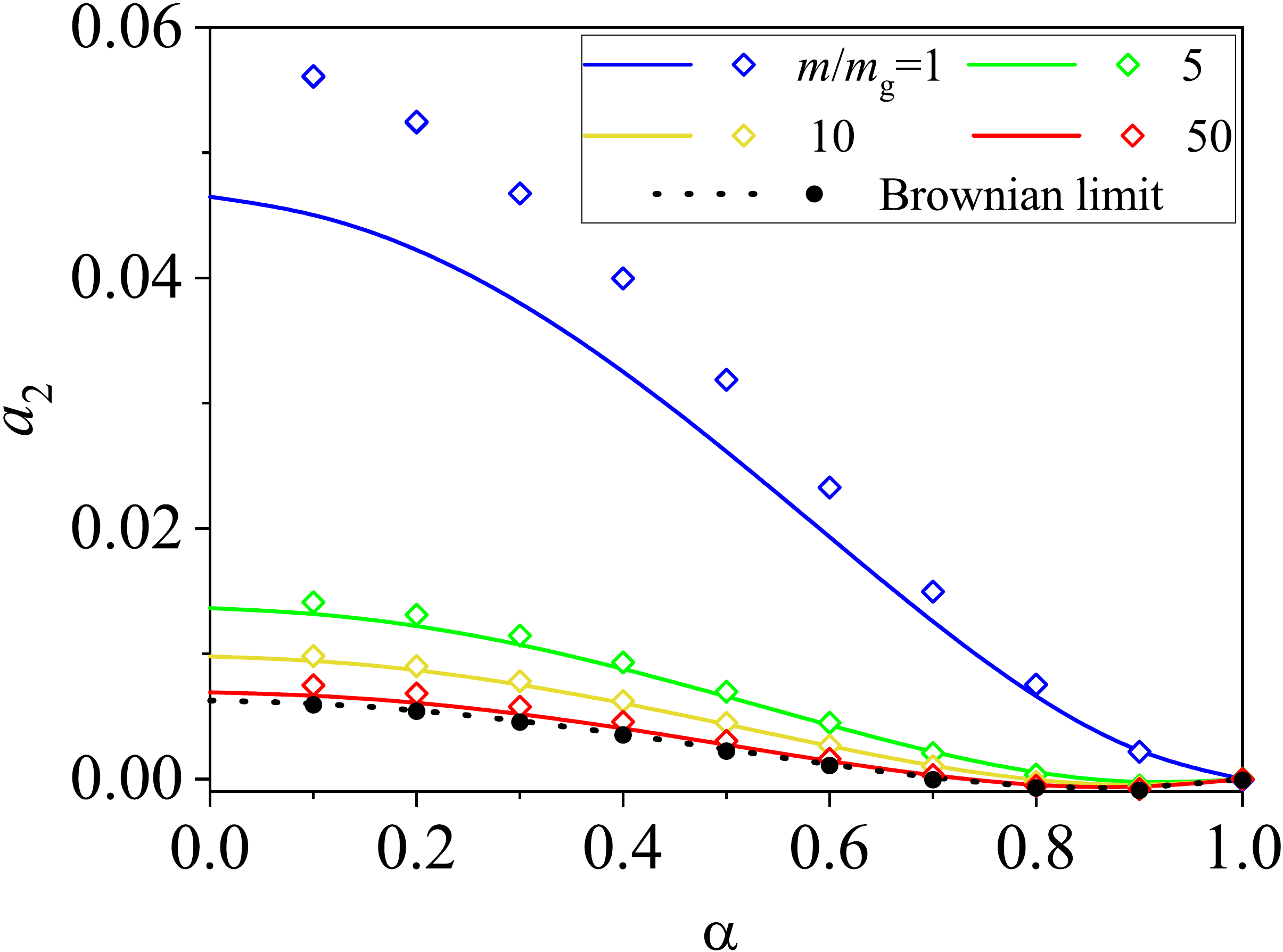}
\end{center}
\caption{Plot of the fourth cumulant $a_2$ as a function of the coefficient of normal restitution $\al$ for $d=3$, $\phi=0.001$, $T_g^*=1000$, and four different values of the mass ratio $m/m_g$ [from top to bottom, $m/m_g=1, 5, 10$, and 50]. The solid lines are the theoretical results obtained from Eq.\ \eqref{2.22} and the symbols are the Monte Carlo simulation results. The dotted line is the result obtained by \cite{GGG19a} by using the Langevin-like suspension model \eqref{1.17} while black circles refer to DSMC simulations implemented using the time-driven approach \eqref{MC4}.
\label{fig2}}
\end{figure}

Figure \ref{fig2} illustrates the $\alpha$-dependence of the cumulant $a_2$ for the same parameters as in Fig.\ \ref{fig1}. As can be seen, the breakdown of energy equipartition makes the system to be in an out-of-equilibrium state where $f\ne f_\text{MB}$. On the other hand, we find that the magnitude of $a_2$ is in general small for not quite large inelasticity (for instance, $\al \gtrsim 0.5$); this result supports the assumption of a low-order truncation (first Sonine approximation) in the polynomial expansion of the distribution function. In addition, the departure of $f$ from its Maxwellian form accentuates when decreasing the mass ratio $m/m_g$ in the same way as the steady granular temperature $T$ moves away from its equilibrium value  $T_g$ reached for elastic collisions ($\alpha=1$). Thus, the magnitude of $a_2$ increases as $m/m_g$ decreases and so, higher-order coefficients in the Sonine approximation could turn out to be significant for strong inelasticity. This could be the reason why we observe some discrepancies between theory and DSMC simulations in Figs.\ \ref{fig1} and \ref{fig2} for $m/m_g=1$, specially in the case of $a_2$. However, these discrepancies are of the same order than those found for dry granular gases \cite[]{mg02}.


\section{Chapman--Enskog expansion. First-order approximation}
\label{sec4}

We perturb now the homogeneous state by small spatial gradients. These perturbations will give nonzero contributions to the pressure tensor and the heat flux vector. The determination of these fluxes will allow us to identify the Navier--Stokes--Fourier transport coefficients of the granular gas. For times longer than the mean free time, we assume that the system evolves towards a \emph{hydrodynamic} regime where the distribution function $f(\mathbf{r}, \mathbf{v};t)$ adopts the form of a \emph{normal} or hydrodynamic solution. This means that all space and time dependence of $f$ only occurs through the hydrodynamic fields $n$, $\mathbf{U}$, and $T$:
\beq
\label{3.1}
f({\bf r},{\bf v},t)=f\left[{\bf v}|n (t), \mathbf{U}(t), T(t) \right].
\end{equation}
The notation on the right hand side indicates a functional dependence on the density, flow velocity and temperature. For low Knudsen numbers (i.e., small spatial variations), the functional dependence \eqref{3.1} can be made local in space by means of an expansion in powers of the gradients $\nabla n$, $\nabla \mathbf{U}$, and $\nabla T$. In this case, $f$ can be expressed in the form
\begin{equation}
f=f^{(0)}+f^{(1)}+f^{(2)}+\cdots,
\label{3.2}
\end{equation}
where the approximation $f^{(k)}$ is of order $k$ in spatial gradients. Here, since we are interested in the Navier--Stokes hydrodynamic equations, only terms up to first order in gradients will be considered in the constitutive equations for the momentum and heat fluxes.

On the other hand, inasmuch as after a transient period and in the absence of spatial gradients, the mean flow velocity $\mathbf{U}$ of the granular gas tends to the mean flow velocity $\mathbf{U}_g$ of the molecular gas, then the velocity difference term $\Delta \mathbf{U}$ must be considered to be at least of first order in the spatial gradients. This implies that the Maxwellian distribution $f_g(\mathbf{v})$ must be also expanded as
\beq
\label{3.3}
f_g(\mathbf{v})=f_g^{(0)}(\mathbf{V})+f_g^{(1)}(\mathbf{V})+\cdots,
\eeq
where
\beq
\label{3.4}
f_g^{(0)}(\mathbf{V})=n_g \Big(\frac{m_g}{2\pi T_g}\Big)^{d/2} \exp \Bigg(-\frac{m_g V^2}{2T_g}\Bigg),
\eeq
and
\beq
\label{3.5}
f_g^{(1)}(\mathbf{V})=-\frac{m_g}{T_g}\mathbf{V}\cdot \Delta \mathbf{U} f_g^{(0)}(\mathbf{V}).
\eeq

According to the expansion \eqref{3.1}, the pressure tensor $P_{ij}$, the heat flux $\mathbf{q}$, and the partial production rates $\zeta$ and $\zeta_g$ must be also expressed accordingly to the perturbation scheme in the forms
\beq
\label{3.6}
P_{ij}=P_{ij}^{(0)}+P_{ij}^{(1)}+\cdots, \quad  \mathbf{q}=\mathbf{q}^{(0)}+\mathbf{q}^{(1)}+\cdots, \quad
\zeta=\zeta^{(0)}+\zeta^{(1)}+\cdots, \quad \zeta_g=\zeta_g^{(0)}+\zeta_g^{(1)}+\cdots
\eeq
In addition, the time derivative $\partial_t$ is also given as
\beq
\label{3.7}
\partial_t=\partial_t^{(0)}+\partial_t^{(1)}+\cdots,
\eeq
where the action of the operators $\partial_t^{(k)}$ on the hydrodynamic fields can be identified when the expansions \eqref{3.6} of the fluxes and the production rates are considered in the macroscopic balance equations \eqref{1.8}--\eqref{1.10}. This is the conventional Chapman--Enskog method \cite[]{CC70,G19} for solving the Boltzmann kinetic equation.

As usual in the Chapman--Enskog method \cite[]{CC70}, the zeroth-order distribution function $f^{(0)}$ defines the hydrodynamic fields $n$, $\mathbf{U}$, and $T$:
\beq
\label{3.7.1}
\left\{n, n \mathbf{U},d n T\right\}=\int d\mathbf{v}\; \left\{1,\mathbf{v}, \frac{m}{2} V^2\right\}f^{(0)}(\mathbf{V}).
\eeq
The requirements \eqref{3.7.1} must be fulfilled at any order in the expansion and so, the distributions $f^{(k)}$ ($k\geq 1$) must thus obey the orthogonality conditions
\beq
\label{3.7.2}
\int d\mathbf{v}\; \left\{1,\mathbf{v}, \frac{m}{2} V^2\right\}f^{(k)}(\mathbf{V})=\left\{0,\mathbf{0},0\right\}.
\eeq
These are the usual solubility conditions of the Chapman--Enskog scheme.

\subsection{Zeroth-order approximation}

To zeroth-order in the expansion, the distribution $f^{(0)}$ verifies the kinetic equation
\beq
\label{3.8}
\partial_t^{(0)}f^{(0)}=J[f^{(0)},f^{(0)}]+J_g[f^{(0)},f_g^{(0)}].
\eeq
The conservation laws at this order give
\beq
\label{3.9}
\partial_t^{(0)}n=0, \quad \partial_t^{(0)}\mathbf{U}=\mathbf{0}, \quad \partial_t^{(0)}T=-T\left(\zeta^{(0)}+\zeta_g^{(0)}\right),
\eeq
where $\zeta^{(0)}$ and $\zeta_g^{(0)}$ are determined from Eqs.\ \eqref{1.13} and \eqref{1.14}, respectively, with the replacements $f \to f^{(0)}$ and $f_g \to f_g^{(0)}$. In particular, as discussed in Sec.\ \ref{sec3}, an accurate approximation to both production rates is given by Eq.\ \eqref{2.8}. In addition, upon obtaining the second relation in Eq.\ \eqref{3.9}, we have accounted for that the distributions $f^{(0)}$ and $f_g^{(0)}$ are isotropic in $\mathbf{V}$ and so, the zeroth-order contribution to the production of momentum vanishes ($\boldsymbol{\mathcal{F}}^{(0)}[f^{(0)}]=\mathbf{0}$).

Since the zeroth-order distribution $f^{(0)}$ qualifies as a normal solution, then $\partial_t^{(0)}f^{(0)}=(\partial_T f^{(0)})(\partial_t^{(0)})$, and Eq.\ \eqref{3.8} can be rewritten as
\beq
\label{3.10}
-\left(\zeta^{(0)}+\zeta_g^{(0)}\right)T\frac{\partial f^{(0)}}{\partial T}=J[f^{(0)},f^{(0)}]+J_g[f^{(0)},f_g^{(0)}].
\eeq
Equation \eqref{3.10} has the same form as the Boltzmann equation \eqref{2.1} for a time-dependent homogeneous state, except that $f^{(0)}(\mathbf{r}, \mathbf{v}; t)$ is the \emph{local} version of the above distribution. Dimensional analysis requires that $f^{(0)}$ has the scaled form
\beq
\label{3.11}
f^{(0)}(\mathbf{r}, \mathbf{v}; t)=n(\mathbf{r};t) v_\text{th}(\mathbf{r};t)^{-d} \varphi\left(\mathbf{c}, T/T_g\right),
\eeq
where $\mathbf{c}=\mathbf{V}/v_\text{th}$, $v_\text{th}(\mathbf{r};t)=\sqrt{2T(\mathbf{r};t)/m}$ being the local thermal velocity. As expected, in contrast to the so-called homogeneous cooling state for (dry) granular gases \cite[]{NE98,G19}, the time dependence of the scaled distribution $\varphi$ does not only occur through the scaled velocity $\mathbf{c}$ but also through the temperature ratio $T/T_g$.

As mentioned before, since $f^{(0)}$ is isotropic in $\mathbf{V}$, the heat flux vanishes ($\mathbf{q}^{(0)}=\mathbf{0}$) and the pressure tensor $P_{ij}^{(0)}=p \delta_{ij}$, where $p=nT$ is the hydrostatic pressure. An estimate to $\zeta^{(0)}$ and $\zeta_g^{(0)}$ in the first-Sonine approximation is provided by Eqs.\ \eqref{2.8}--\eqref{2.10}.

\subsection{First-order approximation}

The determination of the first-order approximation $f^{(1)}(\mathbf{r}, \mathbf{v};t)$ follows similar steps as those made in previous works of granular gases \cite[]{BDKS98,GD99a,GChV13a,GGG19a}. Some technical details involved in the derivation of the kinetic equation verifying $f^{(1)}$ are provided in the Appendix \ref{appA} for the interested reader. To first-order in spatial gradients, the distribution function is given by
\beqa
\label{3.12}
f^{(1)}(\mathbf{V})&=&\boldsymbol{\mathcal{A}}(\mathbf{V})\cdot \nabla \ln T+\boldsymbol{\mathcal{B}}(\mathbf{V})\cdot \nabla \ln n+\mathcal{C}_{ij}\frac{1}{2}
\Big(\partial_i U_j+\partial_j U_i-\frac{2}{d}\delta_{ij}\nabla \cdot \mathbf{U}\Big)\nonumber\\
& & +\mathcal{D}(\mathbf{V}) \nabla \cdot \mathbf{U}+\boldsymbol{\mathcal{E}}\cdot \Delta \mathbf{U},
\eeqa
where the quantities $\boldsymbol{\mathcal{A}}$, $\boldsymbol{\mathcal{B}}$, $\mathcal{C}_{ij}$, $\mathcal{D}$, and $\boldsymbol{\mathcal{E}}$ are the solutions of the following set of coupled linear integral equations:
\beqa
\label{3.13}
& & -\left(\zeta^{(0)}+\zeta_g^{(0)}\right)T\partial_T \mathcal{A}_i-\frac{1}{2}\Bigg[\zeta^{(0)}+\zeta_g^{(0)}\Bigg(1+2\chi \frac{\partial \ln \zeta_g^*}{\partial \chi}\Bigg)\Bigg]\mathcal{A}_i+\mathcal{L}\mathcal{A}_i-\rho^{-1}\frac{\partial f^{(0)}}{\partial V_j}\mathcal{K}_j[\mathcal{A}_i]\nonumber\\
& & -J_g[\mathcal{A}_i,f_g^{(0)}]=A_i,
\eeqa
\beqa
\label{3.14}
& &-\left(\zeta^{(0)}+\zeta_g^{(0)}\right)T\partial_T \mathcal{B}_i+\mathcal{L}\mathcal{B}_i-J_g[\mathcal{B}_i,f_g^{(0)}]-\rho^{-1}\frac{\partial f^{(0)}}{\partial V_j}\mathcal{K}_j[\mathcal{B}_i]=B_i+\Bigg[\zeta^{(0)}+\zeta_g^{(0)}\nonumber\\
& & \times \Bigg(1-\varepsilon \frac{\partial \ln \zeta_g^*}{\partial \varepsilon}\Bigg)\Bigg]\mathcal{A}_i,
\eeqa
\beq
\label{3.15}
-\left(\zeta^{(0)}+\zeta_g^{(0)}\right)T\partial_T \mathcal{C}_{ij}+\mathcal{L}\mathcal{C}_{ij}-\rho^{-1}\frac{\partial f^{(0)}}{\partial V_\ell}\mathcal{K}_\ell[\mathcal{C}_{ij}]-J_g[\mathcal{C}_{ij},f_g^{(0)}]={C}_{ij},
\eeq
\beq
\label{3.16}
-\left(\zeta^{(0)}+\zeta_g^{(0)}\right)T\partial_T \mathcal{D}+\mathcal{L}\mathcal{D}-\left(\zeta_U+\zeta_{Ug}\right)
T\frac{\partial f^{(0)}}{\partial T}-\rho^{-1}\frac{\partial f^{(0)}}{\partial V_i}\mathcal{K}_i[\mathcal{D}]-J_g[\mathcal{D},f_g^{(0)}]=D,
\eeq
\beq
\label{3.17}
-\left(\zeta^{(0)}+\zeta_g^{(0)}\right)T\partial_T \mathcal{E}_i+\mathcal{L}\mathcal{E}_i-\rho^{-1}\frac{\partial f^{(0)}}{\partial V_j}\mathcal{K}_j[\mathcal{E}_i]-J_g[\mathcal{E}_i,f_g^{(0)}]=E_i-\frac{m_g}{T_g}
J_g[f^{(0)},\mathbf{V}f_g^{(0)}].
\eeq
In Eqs.\ \eqref{3.13}--\eqref{3.17}, $\zeta_g^*$ is defined in Eq.\ \eqref{2.7} with the replacement $\zeta_g \to \zeta_g^{(0)}$,
\beq
\label{3.18}
\mathcal{L}X=-\Big(J[f^{(0)},X]+J[X,f^{(0)}]\Big)
\eeq
is the linearized Boltzmann collision operator, and
\beq
\label{3.19}
\mathcal{K}_i[X]=\int d\mathbf{v}\; m V_i J_g[X,f_g^{(0)}].
\eeq
In addition, the coefficients $\mathbf{A}$, $\mathbf{B}$, $C_{ij}$, $D$, and $\mathbf{E}$ are functions of the peculiar velocity $\mathbf{V}$. They are given by
\beq
\label{3.20}
\mathbf{A}(\mathbf{V})=-\mathbf{V}T\frac{\partial f^{(0)}}{\partial T}-\frac{p}{\rho}\frac{\partial f^{(0)}}{\partial \mathbf{V}},
\eeq
\beq
\label{3.21}
\mathbf{B}(\mathbf{V})=-\mathbf{V}n\frac{\partial f^{(0)}}{\partial n}-\frac{p}{\rho}\frac{\partial f^{(0)}}{\partial \mathbf{V}},
\eeq
\beq
\label{3.22}
C_{ij}(\mathbf{V})=V_i \frac{\partial f^{(0)}}{\partial V_j},
\eeq
\beq
\label{3.24}
D(\mathbf{V})=\frac{1}{d}\frac{\partial}{\partial \mathbf{V}}\cdot \left(\mathbf{V}f^{(0)}\right)+\frac{2}{d}
T\frac{\partial f^{(0)}}{\partial T}-f^{(0)}+n\frac{\partial f^{(0)}}{\partial n},
\eeq
\beq
\label{3.25}
\mathbf{E}(\mathbf{V})=-\rho^{-1} \frac{\partial f^{(0)}}{\partial \mathbf{V}}\xi,
\eeq
where
\beq
\label{3.25.1}
\xi=\frac{1}{d}\frac{m_g}{T_g}\int d\mathbf{v}m\mathbf{V}\cdot J_g[f^{(0)},\mathbf{V}f_g^{(0)}].
\eeq
In Eq.\ \eqref{3.16}, we have taken into account that since the production rates $\zeta$ and $\zeta_g$ are scalar quantities, then their first-order corrections in spatial gradients $\zeta^{(1)}$ and $\zeta_g^{(1)}$ must be proportional to $\nabla \cdot \mathbf{U}$ since $\nabla n$, $\nabla T$, and $\Delta \mathbf{U}$ are vectors and the tensor $\partial_i U_j+\partial_j U_i-\frac{2}{d}\delta_{ij}\nabla \cdot \mathbf{U}$ is traceless. Thus,
\beq
\label{3.26}
\zeta^{(1)}=\zeta_U \nabla \cdot \mathbf{U}, \quad \zeta_g^{(1)}=\zeta_{Ug} \nabla \cdot \mathbf{U},
\eeq
where \cite[]{G19}
\beq
\label{3.27}
\zeta_U=\frac{\pi^{(d-1)/2}}{2d\Gamma\Big(\frac{d+3}{2}\Big)}(1-\alpha^2)\frac{m\sigma^{d-1}}{n T}\int d\mathbf{v}_1\int d\mathbf{v}_2 f^{(0)}(\mathbf{V}_1)\mathcal{D}(\mathbf{V}_1) g_{12}^3,
\eeq
\beq
\label{3.28}
\zeta_{Ug}=-\frac{m}{d n T}\int d\mathbf{v} \; V^2\; J_g[\mathcal{D},f_g^{(0)}].
\eeq

The necessary conditions for the solution to the integral equations \eqref{3.13}--\eqref{3.17} to exist [Fredholm alternative \cite[]{MM56}] is that
\beq
\label{3.29}
\int d\mathbf{v}\; \left\{1,\mathbf{v}, \frac{m}{2} V^2\right\}f^{(1)}(\mathbf{V})=\left\{0,\mathbf{0},0\right\}.
\eeq
The conditions \eqref{3.29} on the first-order distribution $f^{(1)}(\mathbf{V})$ are used later to establish the existence of a unique solution of Eqs.\ \eqref{3.13}--\eqref{3.17}. The fulfilment of conditions \eqref{3.29} necessarily requires that the right sides of the integral equations \eqref{3.13}--\eqref{3.17} are orthogonal to the set $(1,\mathbf{v}, \frac{m}{2} V^2)$, namely,
\beq
\label{3.30}
\int d\mathbf{v}\; \left\{1,\mathbf{v}, \frac{m}{2} V^2\right\}
\left(
\begin{array}{c}
\mathbf{A}(\mathbf{V})\\
\mathbf{B}(\mathbf{V})\\
C_{ij}(\mathbf{V})\\
D(\mathbf{V})\\
\mathbf{E}(\mathbf{V})
\end{array}
\right)=
\left(
\begin{array}{c}
0\\
0\\
0\\
0\\
0
\end{array}
\right).
\eeq
It is straightforward to prove fulfilment of the conditions \eqref{3.30} by direct integration using the definitions \eqref{3.20}--\eqref{3.25} of $\mathbf{A}$, $\mathbf{B}$, $C_{ij}$, $D$, and $\mathbf{E}$, respectively.

\subsection{Navier--Stokes transport coefficients}

To first order in spatial gradients and based on symmetry considerations, the pressure tensor $P_{ij}^{(1)}$ and the heat flux $\mathbf{q}^{(1)}$ are given, respectively, by
\beq
\label{3.30.1}
P_{ij}^{(1)}=-\eta \left(\frac{\partial U_i}{\partial r_j}+\frac{\partial U_j}{\partial r_i}-\frac{2}{d}\delta_{ij}\nabla\cdot\mathbf{U}\right),
\eeq
\beq
\label{3.30.2}
\mathbf{q}^{(1)}=-\kappa \nabla T-\overline{\mu} \nabla n-\kappa_U \Delta \mathbf{U}.
\eeq
Here, $\eta$ is the shear viscosity, $\kappa$ is the thermal conductivity, $\overline{\mu}$ is the diffusive heat conductivity, and $\kappa_U$ is the velocity conductivity. To the best of our knowledge, the coefficient $\kappa_U$ is a new transport coefficient for granular suspensions. This coefficient is also present in driven granular mixtures \cite[]{KG13,KG18}. The Navier--Stokes--Fourier transport coefficients are defined as
\beq
\label{3.31}
\eta=-\frac{1}{(d-1)(d+2)}\int d\mathbf{v}\; R_{ij}(\mathbf{V})\mathcal{C}_{ij}(\mathbf{V}),
\eeq
\beq
\label{3.32}
\kappa=-\frac{1}{dT}\int d\mathbf{v}\; \mathbf{S}(\mathbf{V})\cdot \boldsymbol{\mathcal{A}}(\mathbf{V}),
\eeq
\beq
\label{3.33}
\overline{\mu}=-\frac{1}{dn}\int d\mathbf{v}\; \mathbf{S}(\mathbf{V})\cdot \boldsymbol{\mathcal{B}}(\mathbf{V}),
\eeq
\beq
\label{3.33.1}
\kappa_U=-\frac{1}{d}\int d\mathbf{v}\; \mathbf{S}(\mathbf{V})\cdot \boldsymbol{\mathcal{E}}(\mathbf{V}).
\eeq
In Eqs.\ \eqref{3.31}--\eqref{3.33.1}, we have introduced the traceless tensor
\beq
\label{3.33.2}
R_{ij}(\mathbf{V})=m\Big(V_i V_j-\frac{1}{d}V^2 \delta_{ij}\Big),
\eeq
and the vector
\beq
\label{3.34}
\mathbf{S}(\mathbf{V})=\Big(\frac{m}{2}V^2-\frac{d+2}{2}T\Big)\mathbf{V}.
\eeq

\section{Sonine polynomial approximation to the transport coefficients in steady-state conditions}
\label{sec5}

So far, all the results displayed in section \ref{sec4}  for the transport coefficients $\eta$, $\kappa$, $\overline{\mu}$, and $\kappa_U$ are exact. More specifically, their expressions are given by Eqs. \eqref{3.31}--\eqref{3.33.1}, respectively, where the unknowns $\boldsymbol{\mathcal{A}}$, $\boldsymbol{\mathcal{B}}$, $\mathcal{C}_{ij}$, $\mathcal{D}$, and $\boldsymbol{\mathcal{E}}$ are the solutions of the integral equations \eqref{3.13}--\eqref{3.17}, respectively. However, it is easy to see that the solution for general \emph{unsteady} conditions requires to solve numerically a set of coupled differential equations for $\eta$, $\kappa$, $\overline{\mu}$, and $\kappa_U$. Thus, in a desire of achieving analytical expressions of the transport coefficients, we consider steady-state conditions. In this case, the constraint $\zeta^{(0)}+\zeta_g^{(0)}=0$ applies locally and so, the first term of the left-hand side of Eqs.\ \eqref{3.13}--\eqref{3.17} vanish. This yields the set of integral equations
\beq
\label{4.0.1}
-\chi \frac{\partial \ln \zeta_g^*}{\partial \chi}\mathcal{A}_i+\mathcal{L}\mathcal{A}_i-\rho^{-1}\frac{\partial f^{(0)}}{\partial V_j}\mathcal{K}_j[\mathcal{A}_i]-J_g[\mathcal{A}_i,f_g^{(0)}]=A_i,
\eeq
\beq
\label{4.0.2}
\mathcal{L}\mathcal{B}_i-J_g[\mathcal{B}_i,f_g^{(0)}]-\rho^{-1}\frac{\partial f^{(0)}}{\partial V_j}\mathcal{K}_j[\mathcal{B}_i]=B_i-\varepsilon \frac{\partial \ln \zeta_g^*}{\partial \varepsilon}\mathcal{A}_i,
\eeq
\beq
\label{4.0.3}
\mathcal{L}\mathcal{C}_{ij}-\rho^{-1}\frac{\partial f^{(0)}}{\partial V_\ell}\mathcal{K}_\ell[\mathcal{C}_{ij}]-J_g[\mathcal{C}_{ij},f_g^{(0)}]={C}_{ij},
\eeq
\beq
\label{4.0.4}
\mathcal{L}\mathcal{D}-\left(\zeta_U+\zeta_{Ug}\right)
T\frac{\partial f^{(0)}}{\partial T}-\rho^{-1}\frac{\partial f^{(0)}}{\partial V_i}\mathcal{K}_i[\mathcal{D}]-J_g[\mathcal{D},f_g^{(0)}]=D,
\eeq
\beq
\label{4.0.5}
\mathcal{L}\mathcal{E}_i-\rho^{-1}\frac{\partial f^{(0)}}{\partial V_j}\mathcal{K}_j[\mathcal{E}_i]-J_g[\mathcal{E}_i,f_g^{(0)}]=E_i-\frac{m_g}{T_g}
J_g[f^{(0)},\mathbf{V}f_g^{(0)}].
\eeq
Here, all the quantities appearing in Eqs.\ \eqref{4.0.1}--\eqref{4.0.5} are evaluated in the steady state.

Apart from considering steady-state conditions, the determination of the explicit forms of the Navier--Stokes--Fourier transport coefficients requires (i) to solve the set of coupled integral equations \eqref{4.0.1}--\eqref{4.0.5} and additionally, (ii) to know the zeroth-order distribution function $f^{(0)}$. Given that both tasks are extremely intricate, one has to consider some approximations.

Regarding the explicit form of $f^{(0)}$, the results obtained in section \ref{sec3} have shown that the magnitude of the cumulant $a_2$ is in general very small. Therefore, $f^{(0)}(\mathbf{V})$ can be well represented by the Maxwellian distribution, namely,
\beq
\label{4.1}
f^{(0)}(\mathbf{V})\to n \Big(\frac{m}{2\pi T}\Big)^{d/2} \exp \Big(-\frac{m V^2}{2T}\Big).
\eeq
The use of the Maxwellian distribution \eqref{4.1} allows us to get simple but accurate expressions for the Navier--Stokes--Fourier transport coefficients. With the Maxwellian approximation \eqref{4.1}, the collision integral \eqref{3.25.1} can be easily obtained from the results derived by \cite{GM07} for arbitrary coefficients of restitution. Particularising to elastic collisions we get
\beq
\label{4.2}
\xi=\rho\;  \mu\;  \theta^{-1/2}(1+\theta)^{1/2}\gamma,
\eeq
where 
\beq
\label{4.3}
\theta=\frac{m T_g}{m_g T}
\eeq
is the ratio of the mean square velocities of granular and molecular gas particles. The zeroth-contributions to the production rates are $\zeta^{(0)}=(v_\text{th}\zeta^*)/\ell$ and $\zeta_g^{(0)}=(v_\text{th}\zeta_g^*)/\ell$, where
\beq
\label{4.3.1}
\zeta^*=\frac{\sqrt{2}\pi^{(d-1)/2}}{d\Gamma\Big(\frac{d}{2}\Big)}(1-\al^2), \quad \zeta_g^*=2x(1-x^2)\mu^{1/2}\varepsilon,
\eeq
and $x$ and $\varepsilon$ are defined by Eqs.\ \eqref{2.11} and \eqref{2.12}, respectively. The Maxwellian approximation to the steady temperature ratio $T/T_g$ can be obtained by inserting the expressions \eqref{4.3.1} of $\zeta^*$ and $\zeta_g^*$ into the (exact) steady-state condition $\zeta^*+\zeta_g^*=0$. This yields the cubic equation for $x$
\beq
\label{4.3.2}
2x(x^2-1)=\vartheta, \quad \vartheta=\frac{\sqrt{2}\pi^{(d-1)/2}}{d\Gamma\Big(\frac{d}{2}\Big)}\mu^{-1/2}\varepsilon^{-1}(1-\al^2).
\eeq
The physical root of Eq.\ \eqref{4.3.2} can be written as \cite[]{S03a}
\beq
\label{4.3.3}
x=\Bigg\{
\begin{array}{cc}
\frac{\sqrt{3}}{3}\left\{\sqrt{3}\cos\left[\frac{1}{3}\sin^{-1}\left(\frac{3\sqrt{3}}{4}\vartheta\right)\right]+
\sin\left[\frac{1}{3}\sin^{-1}\left(\frac{3\sqrt{3}}{4}\vartheta\right)\right]\right\},&\vartheta \leq \frac{4\sqrt{3}}{9}\\
\frac{2\sqrt{3}}{3}\cosh\left[\frac{1}{3}\cosh^{-1}\left(\frac{3\sqrt{3}}{4}\vartheta\right)\right],&\vartheta \geq \frac{4\sqrt{3}}{9}.
\end{array}
\eeq
According to Eq.\ \eqref{2.11}, the temperature ratio $T/T_g$ in the steady state is then given by
\beq
\label{4.3.4}
\frac{T}{T_g}=\frac{m/m_g}{\left(1+\frac{m}{m_g}\right)x^2-1}.
\eeq

With respect to the functions $\left(\boldsymbol{\mathcal{A}}, \boldsymbol{\mathcal{B}}, \mathcal{C}_{ij}, \mathcal{D}, \boldsymbol{\mathcal{E}}\right)$, it is useful to write them in a series expansion of Sonine (Laguerre) polynomials. In practice only the leading terms in these expansions are retained; they provide a quite accurate description over a wide range of inelasticity. In addition, when the cumulants $a_2$ are neglected, it is straightforward to prove that Eq.\ \eqref{3.24} yields $D=0$ and so, the production rates $\zeta_U=\zeta_{Ug}=0$. Non-vanishing contributions to both production rates (which arise from $a_2$) are expected to be very small \cite[]{GGG19a}. Thus, we will focus here our attention in the Navier--Stokes--Fourier transport coefficients $\eta$, $\kappa$, $\overline{\mu}$, and $\kappa_U$ defined by Eqs.\ \eqref{3.31}--\eqref{3.33.1}, respectively. The procedure for obtaining these transport coefficients is described in the appendix \ref{appB} and only the final expressions in the steady state are provided here.

\subsection{Shear viscosity}

The shear viscosity coefficient $\eta$ is given by
\beq
\label{4.4}
\eta=\frac{\eta_0}{\nu_\eta^*+K'\widetilde{\nu}_\eta\gamma^*},
\eeq
where
\beq
\label{4.5}
\eta_0=\frac{d+2}{8}\frac{\Gamma\Big(\frac{d}{2}\Big)}{\pi^{(d-1)/2}}\sigma^{1-d}\sqrt{m T}
\eeq
is the low density value of the shear viscosity of an ordinary gas of hard spheres ($\al=1$) and
\beq
\label{4.6}
K'=\sqrt{2} \frac{(d+2)\Gamma\Big(\frac{d}{2}\Big)}{8\pi^{(d-1)/2}}.
\eeq
Moreover, we have introduced the (reduced) collision frequencies
\beq
\label{4.7}
\nu_\eta^*=\frac{3}{4d}\Big(1-\al+\frac{2}{3}d\Big)(1+\al),
\eeq
\beqa
\label{4.8}
\widetilde{\nu}_\eta&=&\frac{1}{(d-1)(d+2)}\Big(\frac{m}{m_g}\Big)^3 \mu_g \Big(\frac{T_g}{T}\Big)^2\theta^{-1/2}
\Bigg\{2(d+3)(d-1)\left(\mu-\mu_g \theta\right)\theta^{-2}(1+\theta)^{-1/2}\nonumber\\
& & +2d(d-1)\mu_g \theta^{-2}(1+\theta)^{1/2}+2(d+2)(d-1)\theta^{-1}(1+\theta)^{-1/2}\Bigg\},
\eeqa
where $\gamma^*$ is defined in Eq.\ \eqref{2.12}. It is important to recall that all the quantities appearing in Eq.\ \eqref{4.4} are evaluated at the steady-state conditions.

\subsection{Thermal conductivity, diffusive heat conductivity, and velocity conductivity}

We consider here the transport coefficients associated with the heat flux. The thermal conductivity coefficient $\kappa$ is
\beq
\label{4.9}
\kappa=\frac{d-1}{d}\frac{\kappa_0}{\nu_\kappa^*+K'\left(\widetilde{\nu}_\kappa+\beta\right)\gamma^*},
\eeq
where
\beq
\label{4.10}
\kappa_0=\frac{d(d+2)}{2(d-1)}\frac{\eta_0}{m}
\eeq
is the low density value of the thermal conductivity for an ordinary gas of hard spheres and
\beq
\label{4.11}
\beta=\left(x^{-1}-3x\right)\mu^{3/2}\left(\frac{T_g}{T}\right)^{1/2}.
\eeq
In Eq.\ \eqref{4.9}, we have introduced the (reduced) collision frequencies
\beq
\label{4.12}
\nu_\kappa^*=\frac{1+\al}{d}\left[\frac{d-1}{2}+\frac{3}{16}(d+8)(1-\al)\right](1+\al),
\eeq
\beq
\label{4.13}
\widetilde{\nu}_\kappa=\frac{1}{2(d+2)}\mu \frac{\theta}{1+\theta} \Big[G-(d+2)\frac{1+\theta}{\theta}F\Big],
\eeq
where
\beqa
\label{4.14}
F&=&(d+2)(2 \delta +1)+4(d-1)\mu_g \delta \theta^{-1}(1+\theta)+3(d+3)\delta^2\theta^{-1}+(d+3)\mu_g^2 \theta^{-1}(1+\theta)^2\nonumber\\
& & -(d+2)\theta^{-1}(1+\theta),
\eeqa
\beqa
\label{4.15}
G&=&(d+3)\mu_g^2 \theta^{-2}(1+\theta)^2\left[d+5+(d+2)\theta\right]-\mu_g(1+\theta)\Big\{4(1-d)\delta \theta^{-2}\left[d+5+(d+2)\theta\right]
\nonumber\\
& & -8(d-1)\theta^{-1}\Big\}+3(d+3)\delta^2\theta^{-2}\left[d+5+(d+2)\theta\right] +2\delta\theta^{-1}\left[24+11d+d^2\right.\nonumber\\
& & \left.+(d+2)^2\theta\right]+(d+2)\theta^{-1} \left[d+3+(d+8)\theta\right]-(d+2)\theta^{-2}(1+\theta)\left[d+3+(d+2)\theta\right].
\nonumber\\
\eeqa
Here, $\delta\equiv \mu-\mu_g \theta$.

The diffusive heat conductivity $\overline{\mu}$ can be written as
\beq
\label{4.16}
\overline{\mu}=\frac{K' T}{n}\frac{\kappa \zeta^*}{\nu_\kappa^*+K'\widetilde{\nu}_\kappa \gamma^*}.
\eeq
Finally, the velocity conductivity $\kappa_U$ is given by
\beq
\label{4.17}
\kappa_U=-\frac{n T}{2}\frac{K' \mu (1+\theta)^{-1/2}\theta^{-1/2}H}{\nu_\kappa^*+K'\widetilde{\nu}_\kappa \gamma^*}\gamma^*,
\eeq
where
\beq
\label{4.18}
H=(d+2)(1+2\delta)+4(1-d)\mu_g (1+\theta) \delta-3(d+3)\delta^2-(d+3)\mu_g^2 (1+\theta)^2.
\eeq

\subsection{Brownian limit}

Equations \eqref{4.4}, \eqref{4.9}, \eqref{4.16}, and \eqref{4.17} provide the expressions of the transport coefficients $\eta$, $\kappa$, $\overline{\mu}$, and $\kappa_U$, respectively, for arbitrary values of the mass ratio $m/m_g$. As did before in the homogeneous state, it is quite interesting to consider the limiting case $m/m_g\to \infty$ (Brownian limit). In this limit case, $\mu_g\to 0$, $\mu\to 1$, $\chi \equiv \text{finite}$, and so $\theta\to 0$, $x\to \chi^{-1/2}$, $\delta\to 1-\chi^{-1}$, and $\beta\to 1-3\chi^{-1}$. This yields the results $\widetilde{\nu}_\eta\to 2$ and $\widetilde{\nu}_\kappa\to 3$, so that in the Brownian limit Eqs.\ \eqref{4.4}, \eqref{4.9}, \eqref{4.16}, and \eqref{4.17} reduce to
\beq
\label{4.19}
\eta\to \frac{\eta_0}{\nu_\eta^*+2K'\gamma^*}, \quad \kappa\to \frac{d-1}{d}\frac{\kappa_0}{\nu_\kappa^*+K'\left(\gamma^*-\frac{3}{2}\zeta^*\right)\gamma^*},
\eeq
\beq
\label{4.20}
\overline{\mu}\to \frac{\kappa T}{n}\frac{K'\zeta^*}{\nu_\kappa^*+3K' \gamma^*}, \quad \kappa_U \to 0.
\eeq
Equations \eqref{4.19} and \eqref{4.20} agree with the results obtained by \cite{GGG19a} by using the suspension model \eqref{1.17}. This confirms the self-consistency of the results obtained in this paper for general values of the mass ratio.

\subsection{Some illustrative systems}

\begin{figure}
\begin{center}
\includegraphics[width=.55\columnwidth]{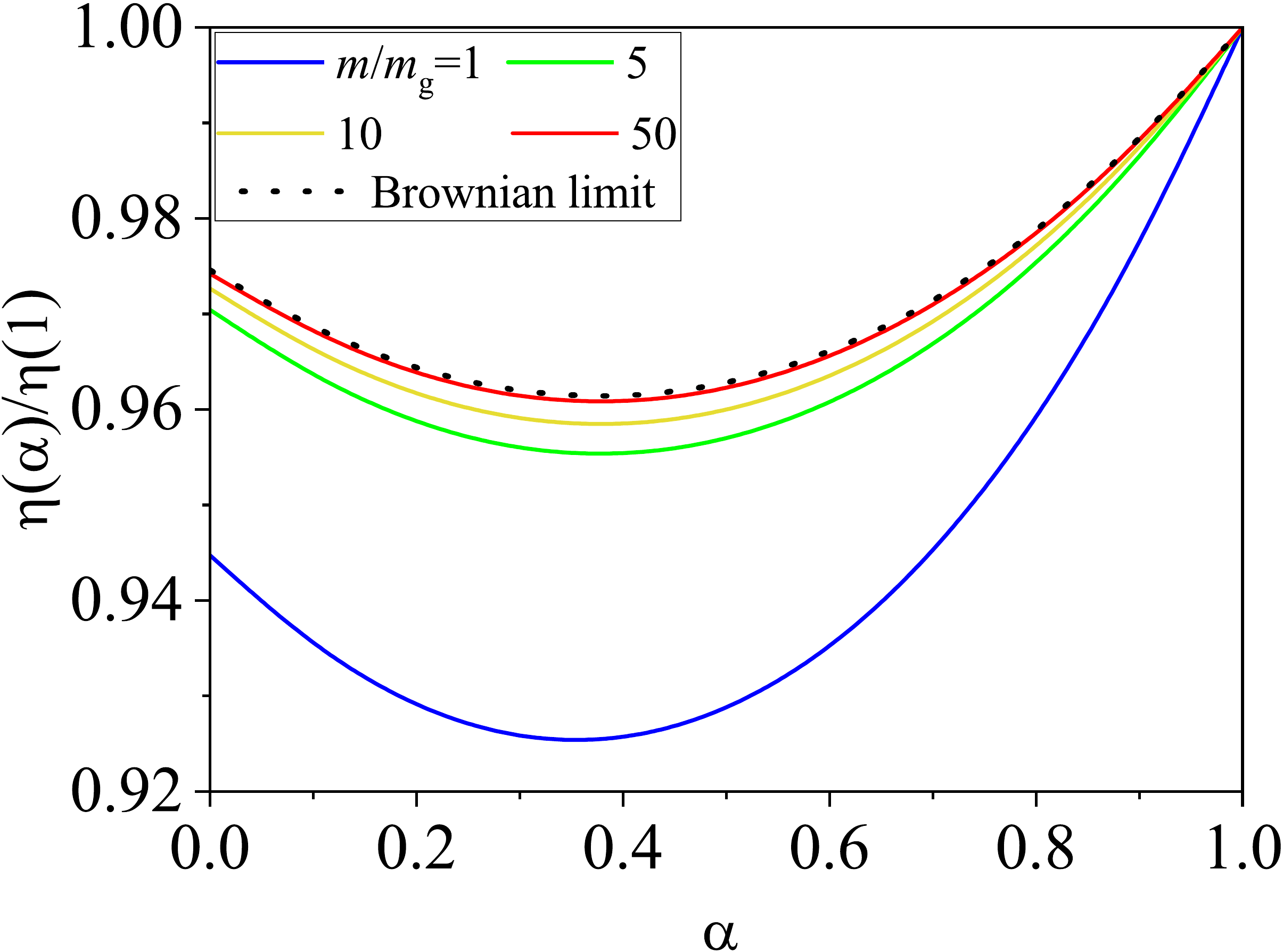}
\end{center}
\caption{Plot of the (scaled) shear viscosity coefficient $\eta(\al)/\eta(1)$ versus the coefficient of normal restitution $\al$ for $d=3$,
$\phi=0.001$, $T_g^*=1000$, and four different values of the mass ratio $m/m_g$ [from top to bottom, $m/m_g=50, 10, 5$, and 1]. The solid lines are the results derived in this paper while the dotted line is the result obtained by \cite{GGG19a} by using the suspension model \eqref{1.17}. Here, $\eta(1)$ refers to the shear viscosity coefficient when collisions between grains are elastic ($\al=1$).
\label{fig3}}
\end{figure}
\begin{figure}
\begin{center}
\includegraphics[width=.55\columnwidth]{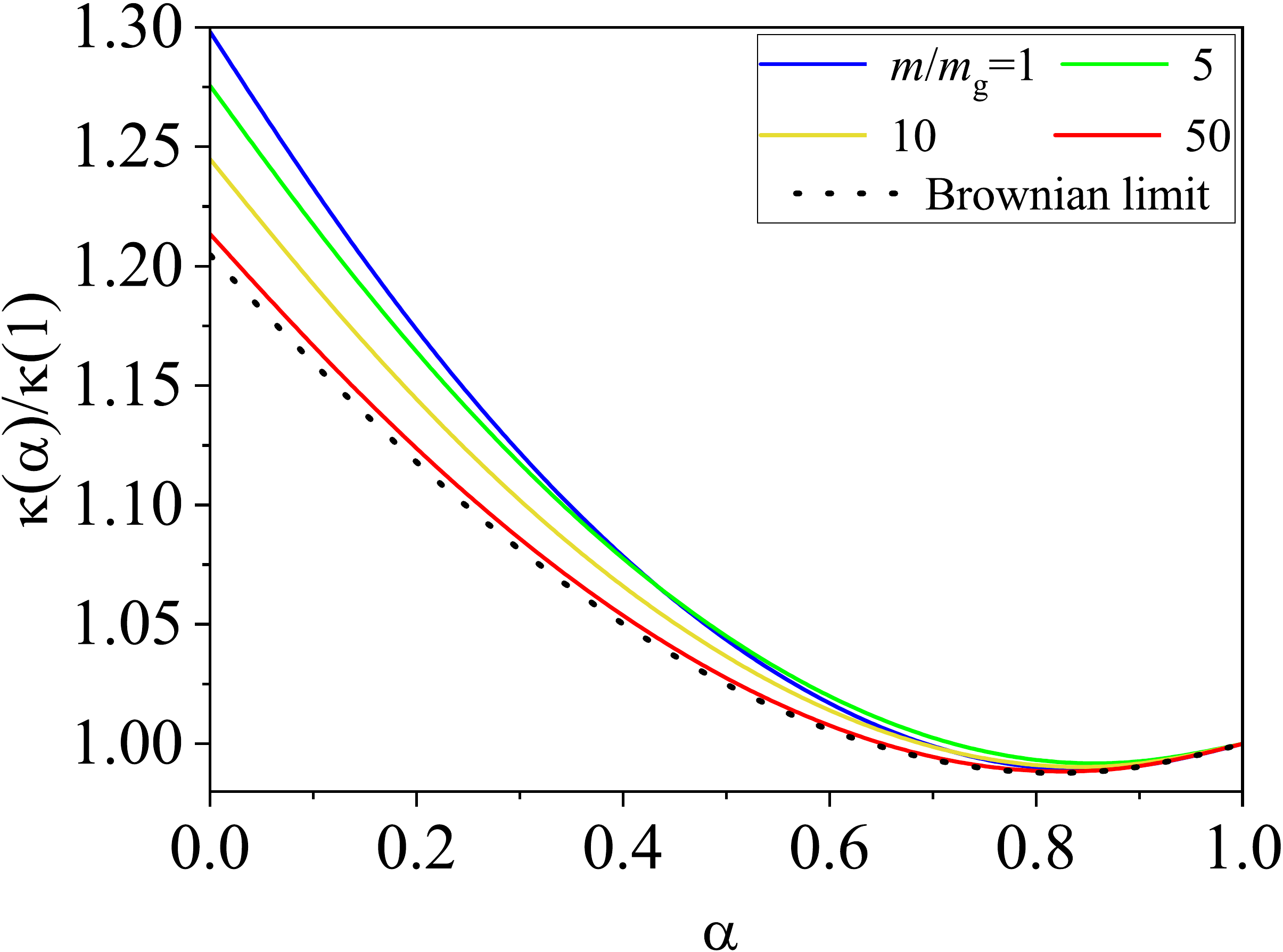}
\end{center}
\caption{Plot of the (scaled) thermal conductivity coefficient $\kappa(\al)/\kappa(1)$ versus the coefficient of normal restitution $\al$ for $d=3$, $\phi=0.001$, $T_g^*=1000$, and four different values of the mass ratio $m/m_g$ [from top to bottom, $m/m_g=1, 5, 10$, and 50]. The solid lines are the results derived in this paper while the dotted line is the result obtained by \cite{GGG19a} by using the suspension model \eqref{1.17}. Here, $\kappa(1)$ refers to the thermal conductivity coefficient when collisions between grains are elastic ($\al=1$).
\label{fig4}}
\end{figure}
\begin{figure}
\begin{center}
\includegraphics[width=.55\columnwidth]{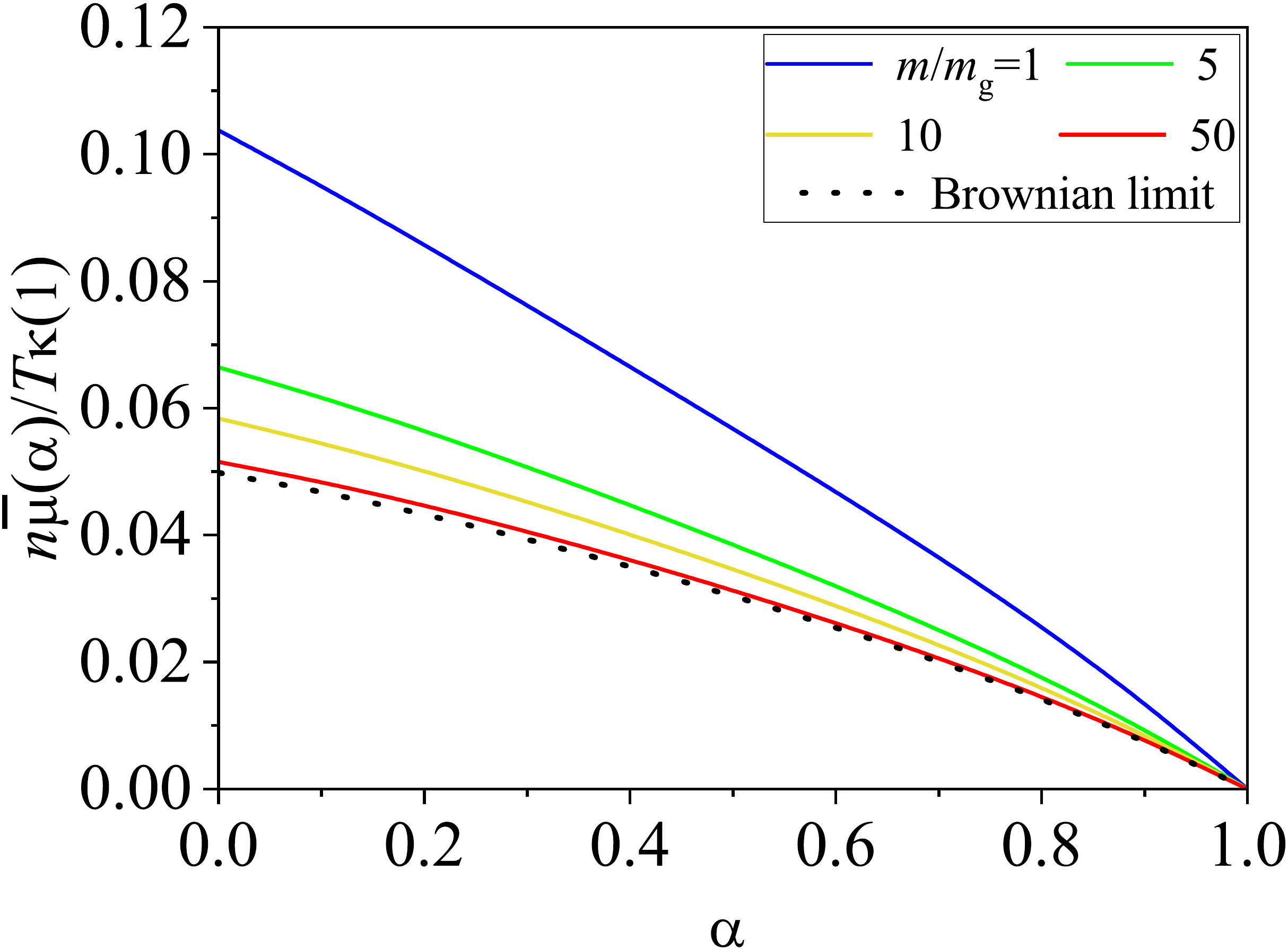}
\end{center}
\caption{Plot of the (scaled) diffusive heat conductivity coefficient $n\overline{\mu}(\al)/T\kappa(1)$ versus the coefficient of normal restitution $\al$ for $d=3$, $\phi=0.001$, $T_g^*=1000$, and four different values of the mass ratio $m/m_g$ [from top to bottom, $m/m_g=1, 5, 10$, and 50]. The solid lines are the results derived in this paper while the dotted line is the result obtained by \cite{GGG19a} by using the suspension model \eqref{1.17}. Here, $\kappa(1)$ refers to the thermal conductivity coefficient when collisions between grains are elastic ($\al=1$).
\label{fig5}}
\end{figure}
\begin{figure}
\begin{center}
\includegraphics[width=.55\columnwidth]{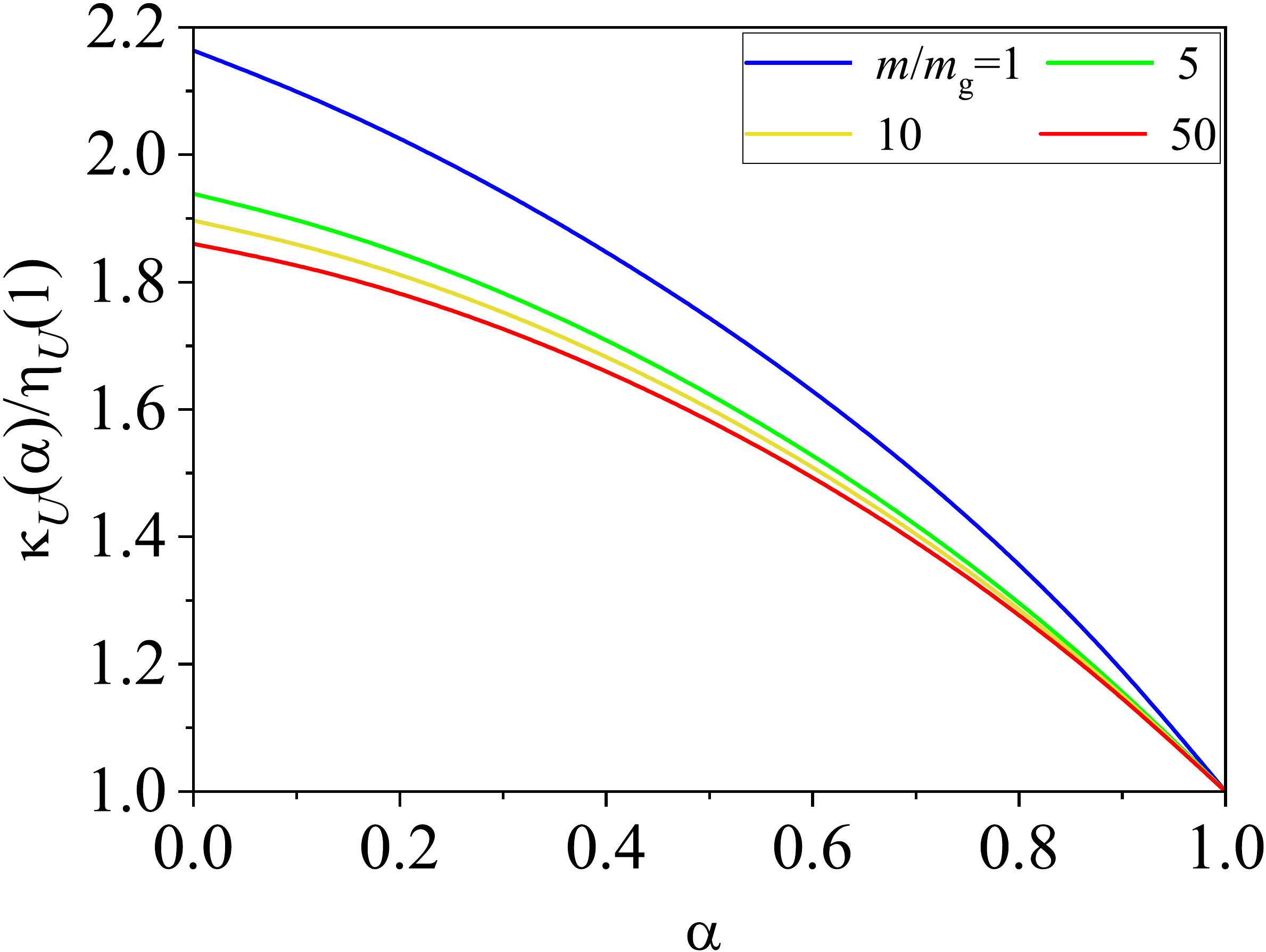}
\end{center}
\caption{Plot of the (scaled) velocity conductivity coefficient $\kappa_U(\al)/\kappa_U(1)$ versus the coefficient of normal restitution $\al$ for $d=3$, $\phi=0.001$, $T_g^*=1000$, and four different values of the mass ratio $m/m_g$ [from top to bottom, $m/m_g=1, 5, 10$, and 50]. Here, $\kappa_U(1)$ refers to the velocity conductivity coefficient when collisions between grains are elastic ($\al=1$). Note that in the Brownian limit ($m/m_g\to \infty$), $\kappa_U\to 0$, in agreement with the result obtained by \cite{GGG19a} by using the suspension model \eqref{1.17}.
\label{fig6}}
\end{figure}

In the steady state, the expressions of the Navier--Stokes--Fourier transport coefficients $\eta$, $\kappa$, $\overline{\mu}$, and $\kappa_U$ are provided by Eqs.\ \eqref{4.4}, \eqref{4.9}, \eqref{4.16}, and \eqref{4.17}, respectively. As in previous works on transport in granular gases \cite[]{BDKS98,GD99a,GTSH12,GGG19a}, to highlight the $\al$-dependence of the transport coefficients, they are scaled with respect to their values for elastic collisions. This scaling cannot be made in the case of the diffusive heat conductivity $\overline{\mu}$ since this coefficient vanishes for $\al=1$. In this case, we consider the scaled coefficient $n\overline{\mu}/T\kappa(1)$, where $\kappa(1)$ refers to the value of the thermal conductivity \eqref{4.9} for elastic collisions. All these scaled coefficients exhibit a complex dependence on the coefficient of restitution $\al$, the mass ratio $m/m_g$, the volume fraction $\phi$ [through the parameter $\varepsilon$ defined by Eq.\ \eqref{2.12}], and the reduced temperature $T_g^*$ of the molecular gas. Moreover, these dimensionless transport coefficients are defined in terms of the temperature ratio $T/T_g$, which is given by Eqs.\ \eqref{4.3.3} and \eqref{4.3.4}.

Figures \ref{fig3}--\ref{fig6} show $\eta(\al)/\eta(1)$, $\kappa(\al)/\kappa(1)$, $n\overline{\mu}/T\kappa(1)$, and $\kappa_U(\al)/\kappa_U(1)$, respectively, as functions of the coefficient of restitution $\al$. Here, $\eta(1)$ and $\kappa_U(1)$ correspond to the values of $\eta$ and $\kappa_U$ for elastic collisions. Moreover, in the above plots we consider a three-dimensional system ($d=3$) with $\phi=0.001$ (very dilute granular gas), $T_g^*=1000$, and four different values of the mass ratio: $m/m_g=1, 5, 10,$ and 50. We have also plotted the results obtained by \cite{GGG19a} by using the suspension model \eqref{1.17}. This model is expected to apply in the Brownian limit ($m \gg m_g$).

We observe that the deviations of the transport coefficients from their elastic forms are in general significant, specially when $m=m_g$. While the (scaled) shear viscosity and thermal conductivity coefficients exhibit a non-monotonic dependence on inelasticity, the (scaled) heat diffusive and velocity conductivity coefficients increase with increasing inelasticity, regardless of the value of the mass ratio considered. In addition, while $\eta(\al)<\eta(1)$, the opposite happens for the thermal conductivity since $\kappa(\al)>\kappa(1)$. With respect to the dependence on the mass ratio $m/m_g$, at a fixed value of the coefficient of restitution, it is quite apparent that while the (scaled) shear viscosity increases with increasing the mass ratio, the (scaled) thermal conductivity decreases with increasing the mass ratio. The same happens for the (scaled) coefficients $n\overline{\mu}/T\kappa(1)$ and $\kappa_U(\al)/\kappa_U(1)$ since both scaled coefficients decrease as the mass ratio increases. We also see that in the case $m/m_g=50$, the results derived here for $\eta(\al)/\eta(1)$, $\kappa(\al)/\kappa(1)$, and $n\overline{\mu}/T\kappa(1)$ practically coincide with those obtained in the Brownian limit  by \cite{GGG19a}. However, in the case $m/m_g=50$, the (scaled) velocity conductivity coefficient $\kappa_U(\al)/\kappa_U(1)$ (which vanishes in the Brownian limit) is still clearly different from zero.

Although the results obtained here for the (scaled) transport coefficients depend on the values of the mass ratio and the (reduced) temperature of the molecular gas, it is worthwhile comparing the present results with those obtained for \emph{dry} granular gases (i.e., in the absence of the molecular gas). In the case of the shear viscosity, a comparison between both systems (with and without the gas phase) shows significant discrepancies [see for instance, Fig.\ 3.1 of the textbook of \cite{G19}] even at a qualitative level since while $\eta$ increases with inelasticity for dry granular gases, the opposite happens here whatever the mass ratio considered. On the other hand, a more qualitative agreement is found for the the thermal conductivity [see for instance, Fig.\ 3.2 of \cite{G19}] since $\kappa$ increases with decreasing $\al$ in both systems. In any case, important quantitative differences appear at strong dissipation since the influence of inelasticity on $\kappa$ is more relevant in the dry case than in the presence of the molecular gas. A similar conclusion is reached for the heat diffusive coefficient $\overline{\mu}$ [see for instance, Fig.\ 3.3 of the textbook of \cite{G19}] where the magnitude of this (scaled) coefficient for dry granular gases is much more large than the one found here for granular suspensions. In fact, when the particles of the granular gas are much more heavier than those of the molecular gas, given that the magnitude of $n\overline{\mu}/T\kappa(1)$ is much smaller than that of $\kappa(\al)/\kappa(1)$ then, one could neglect the contribution coming from the density gradient in the heat flux and assume the validity of Fourier's law $\mathbf{q}=-\kappa \nabla T$.

\section{Linear stability analysis of the homogeneous steady state}
\label{sec6}

Once the transport coefficients of the granular gas are known, the corresponding Navier--Stokes hydrodynamic equations can be explicitly displayed. To derive them, one has to take into account first that the the production of momentum $\boldsymbol{\mathcal F}[f]$ to first order in spatial gradients can be written as
\beq
\label{6.1}
\boldsymbol{\mathcal F}^{(1)}[f^{(1)}]=-\xi \Delta \mathbf{U}+\frac{\rho}{d+2}\mu\gamma\left(\frac{\kappa}{n}\frac{\partial \ln T}{\partial r_i}+\frac{\overline{\mu}}{T}\frac{\partial \ln n}{\partial r_i}+\frac{\kappa_U}{nT}\Delta U_i\right)X,
\eeq
where $\xi$ is given by Eq.\ \eqref{4.2} and
\beq
\label{6.2}
X(\theta)=\theta^{-1/2}\left(1+\theta\right)^{-1/2}.
\eeq
Thus, when the constitutive equations \eqref{3.30.1}--\eqref{3.30.2} and Eq.\ \eqref{6.1} are substituted into the (exact) balance equations \eqref{1.8}--\eqref{1.10}, one gets the Navier--Stokes hydrodynamic equations for a granular gas immersed in a molecular gas:
\begin{equation}
\label{6.3}
D_t n+n\nabla\cdot\mathbf{U}=0,
\end{equation}
\beqa
\label{6.4}
\rho D_t U_i+\frac{\partial p}{\partial r_i}&=&\frac{\partial}{\partial r_j}\left[\eta\left(\frac{\partial U_j}{\partial r_i}+\frac{\partial U_i}{\partial r_j}-\frac{2}{d}\delta_{ij}
\nabla\cdot\mathbf{U}\right)\right]-\xi \Delta U_i\nonumber\\
& & +\frac{\rho}{d+2}\mu\gamma X\left(\frac{\kappa}{n}\frac{\partial \ln T}{\partial r_i}+\frac{\overline{\mu}}{T}\frac{\partial \ln n}{\partial r_i}+\frac{\kappa_U}{nT}\Delta U_i\right),
\eeqa
\beqa
\label{6.5}
D_tT+T\Big(\zeta^{(0)}+\zeta_g^{(0)}\Big)&=&\frac{2}{dn}\nabla\cdot\left(\kappa\nabla T+\overline{\mu}\nabla n+\kappa_U \Delta \mathbf{U}\right)+\frac{2}{dn}\Big[\eta\Big(\frac{\partial U_j}{\partial r_i}+\frac{\partial U_i}{\partial r_j}\nonumber\\
& & -\frac{2}{d}\delta_{ij}
\nabla\cdot\mathbf{U}\Big)\frac{\partial U_i}{\partial r_j}-\frac{2}{d}T\nabla\cdot\mathbf{U}\Big].
\eeqa
As said in section \ref{sec5}, we have not considered in Eq.\ \eqref{6.5} the first-order contributions to $\zeta$ and $\zeta_g$ since they vanish when non-Gaussian corrections to the distribution function $f^{(0)}$ are neglected. In addition, as already mentioned in several previous works \cite[]{G05,GMD06}, the above production rates should also include second-order contributions in spatial gradients. However, in the case of a dry dilute granular gas \cite[]{BDKS98}, it has been shown that these contributions are very small and hence, they can be neglected in the hydrodynamic equations. We expect here that the same happens for a granular suspension. Apart from the above approximations, the Navier--Stokes hydrodynamic equations \eqref{6.3}--\eqref{6.5} are exact to second order in the spatial gradients of $n$, $\mathbf{U}$, and $T$.

A simple solution of Eqs.\ \eqref{6.3}--\eqref{6.5} corresponds to the HSS studied in section \ref{sec2}. A natural question is if actually the HSS may be unstable with respect to long enough wavelength perturbations, as occurs for \emph{dry} granular fluids in freely cooling flows \cite[]{GZ93,M93}. This is one of the most characteristic features of granular gases; its origin is associated with the inelasticity of collisions. On the other hand, the stability of the HSS was also analysed by \cite{GGG19a} in the Brownian limit case. The results show that the HSS is always linearly stable, in contrast to what happens for dry granular fluids. Since the present work generalises the study carried out by \cite{GGG19a} to arbitrary values of the mass ratio $m/m_g$, it is worth to check out if the previous theoretical results \cite[]{GGG19a} are indicative of what occurs when the mass-ratio dependence of the transport coefficients is accounted for. This is the main objective of this section.

As usual, we assume that we slightly perturb the HSS by small spatial gradients and hence, the Navier--Stokes hydrodynamic equations \eqref{6.3}--\eqref{6.5} are linearised around the HSS. This state describes a homogeneous state ($\nabla n_H=\nabla T_H=0$) with vanishing flow velocity fields ($\mathbf{U}=\mathbf{U}_g=\mathbf{0}$). In addition, the steady condition is $\zeta_H^{(0)}+\zeta_{gH}^{(0)}=0$. Here, the subscript $H$ denotes quantities evaluated in the HSS. We suppose that the deviations
\beq
\label{6.6}
\delta y_{\beta}(\mathbf{r},t)=y_{\beta}(\mathbf{r},t)-y_{\beta,H}
\eeq
are small. Here, $\delta y_{\beta}(\mathbf{r},t)$ denotes the deviations of the hydrodynamic fields
\beq
\label{6.6.1}
\left\{y_\beta; \beta=1,\cdots, d+2\right\}\equiv \left\{n, \mathbf{U}, T\right\}
\eeq
from their values in the homogeneous \textit{steady} state. Moreover, as usual in the simulations of clustering instabilities in fluid-solid systems \cite[]{FLYH17}, the molecular gas properties are assumed to be constant and so, they are not perturbed.

Although the reference HSS is stationary [and so, in contrast to what happens in dry granular gases, one does not have to eliminate the time dependence of the transport coefficients through adequate changes of space and time \cite[]{BDKS98,G05}], in order to compare the present stability analysis with the one carried out in the Brownian limit \cite[]{GGG19a}, we introduce the following space and time variables:
\beq
\label{6.7}
\tau=\frac{v_0}{2\ell}t, \quad \mathbf{r}'=\frac{\mathbf{r}}{2\ell},
\eeq
where $v_0=\sqrt{T_{H}/m}$ and $\ell=1/(n_H\sigma^{d-1})$. The dimensionless time scale $\tau$ measures the average number of collisions per particle in the time interval between 0 and $t$. The unit length $\mathbf{r}'$ is proportional to the mean free path $\ell$ of solid particles.

The resulting equations for $\delta n$, $\delta \mathbf{U}$, and $\delta T$ can be easily obtained when one substitutes the ansatz \eqref{6.6} into Eqs.\ \eqref{6.3}--\eqref{6.5} and neglects terms of second and higher order in the perturbations. After some algebra, one gets the set of differential equations:
\beq
\label{6.8}
\frac{\partial}{\partial \tau}\frac{\delta n}{n_H}+\nabla'\cdot \frac{\delta \mathbf{U}}{v_0}=0,
\eeq
\beqa
\label{6.9}
\frac{\partial}{\partial \tau}\frac{\delta \mathbf{U}}{v_0}+\nabla'\Big(\frac{\delta n}{n_H}+\frac{\delta T}{T_H}\Big)&=& \frac{d-2}{2d}\eta^*\nabla'\nabla'\cdot \frac{\delta \mathbf{U}}{v_0}+\frac{1}{2}\eta^*\nabla^{'2}\frac{\delta \mathbf{U}}{v_0}-2\xi^* \frac{\delta \mathbf{U}}{v_0}\nonumber\\ & &+\frac{\sqrt{2}d}{d+2}\mu\gamma^*\left(D_T^*\nabla'\frac{\delta T}{T_H}+\overline{\mu}^*\nabla'\frac{\delta n}{n_H}+2\kappa_U^*\frac{\delta\mathbf{U}}{v_0}\right)X,
\nonumber\\
\eeqa
\beq
\label{6.10}
\frac{\partial}{\partial \tau}\frac{\delta T}{T_H}+2\sqrt{2}\zeta^*\Big(\frac{\delta n}{n_H}+\frac{1}{2}\frac{\delta T}{T_H}\Big)+2\sqrt{2}\gamma^*\overline{\zeta}_g \frac{\delta T}{T_H}=D_T^* \nabla^{'2}\frac{\delta T}{T_H}+\overline{\mu}^*\nabla^{'2}\frac{\delta n}{n_H}-\left(\frac{2}{d}-2\kappa_U^*\right)\nabla'\cdot \frac{\delta \mathbf{U}}{v_0},
\eeq
where $\nabla'\equiv \partial/\partial \mathbf{r}'$ and we have introduced the reduced quantities
\beq
\label{6.11}
\eta^*=\frac{\eta_H}{\sigma^{1-d}\sqrt{mT_H}}, \quad \xi^*=\frac{\ell \xi_H}{\rho_{H} v_0}, \quad D_T^*=\frac{\kappa_H}{d\sigma^{1-d}\sqrt{T_H/m}}, \quad \overline{\mu}^*=\frac{\rho_H\overline{\mu}_H}{d\sigma^{1-d}T_H\sqrt{mT_H}},
\eeq
\beq
\label{6.12}
\kappa_U^*=\frac{\kappa_{UH}}{d n_H T_H}, \quad\overline{\zeta}_{g}=\Bigg(\frac{\mu T_H}{T_g}\Bigg)^{1/2}\Bigg[x_H\left(1-x_H^2\right)
-\frac{\mu T_g}{x_H T_H}\left(1-3x_H^2\right)\Bigg].
\eeq
Here, $x_H$ is defined by Eq.\ \eqref{2.11} with the replacement $T\to T_H$.

Then, a set of Fourier transformed dimensionless variables are introduced as
\begin{equation}
\label{6.13}
\rho_{\mathbf{k}}(\tau)=\frac{\delta n_{\mathbf{k}}(\tau)}{n_H},\quad \mathbf{w}_{\mathbf{k}}(\tau)=\frac{\delta\mathbf{U}_{\mathbf{k}}(\tau)}{v_0},
\quad\theta_{\mathbf{k}}(\tau)=\frac{\delta T_{\mathbf{k}}(\tau)}{T_H},
\end{equation}
where the elements of the set $\delta y_{\mathbf{k}\beta}\equiv\left\{\rho_{\mathbf{k}}(\tau),\mathbf{w}_{\mathbf{k}}(\tau),\theta_{\mathbf{k}}(\tau)\right\}$ are defined as
\beq
\label{6.14}
\delta y_{\mathbf{k}\beta}(\tau)=\int d\mathbf{r}'\text{e}^{-i\mathbf{k}\cdot\mathbf{r}'}\delta y_{\beta}(\mathbf{r}',\tau).
\eeq
Note that here the wave vector $\mathbf{k}$ is dimensionless.

As expected \cite[]{BDKS98,G05}, the $d-1$ transverse velocity components
$\mathbf{w}_{\mathbf{k}\perp}=\mathbf{w}_{\mathbf{k}}-\left(\mathbf{w}_{\mathbf{k}}\cdot\widehat{\mathbf{k}}\right)\widehat{\mathbf{k}}$ (orthogonal to the wave vector $\mathbf{k}$) decouple from the other three modes. Their evolution equation is
\begin{equation}
\label{6.15}
\frac{\partial \mathbf{w}_{\mathbf{k}\perp}}{\partial\tau}+\left(\frac{1}{2}\eta^*k^2+2\xi^*
-\frac{2\sqrt{2}d}{d+2}\mu\gamma^*\kappa_U^*X\right)\mathbf{w}_{\mathbf{k}\perp}=0.
\end{equation}
In the Brownian limit ($m/m_g\to \infty$), $\xi^*=\sqrt{2}\gamma^*$, $X\to 0$, and Eq.\ \eqref{6.15} is consistent with the one obtained in previous works \cite[]{GGG19a}. The solution to Eq.\ \eqref{6.15} is
\begin{equation}
\label{6.16}
\mathbf{w}_{\mathbf{k}\perp}(\mathbf{k},\tau)=\mathbf{w}_{\mathbf{k}\perp}(0)\exp\left[\lambda_\perp(k)\tau\right], \quad
\lambda_\perp(k)=\frac{2\sqrt{2}d}{d+2}\mu\gamma^*\kappa_U^*X-2\xi^*-\frac{1}{2}\eta^*k^2.
\end{equation}
A systematic analysis of the dependence of $\lambda_\perp(k)$ on the parameter space of the system shows that $\lambda_\perp(k)$ is always \emph{negative} and hence, the transversal shear modes $\mathbf{w}_{\mathbf{k}\perp}(\tau)$ are linearly \emph{stable}.

The analysis of the remaining three longitudinal modes ($\rho_\mathbf{k}$, $\theta_\mathbf{k}$, and the longitudinal velocity component of the velocity field, $w_{\mathbf{k}\parallel}=\mathbf{w}_{\mathbf{k}}\cdot\widehat{\mathbf{k}}$) is more intricate since these modes are coupled. In matrix form, they verify the equation
\begin{equation}
\label{6.17}
\frac{\partial\delta y_{\mathbf{k}\beta}(\tau)}{\partial\tau}+M_{\beta\mu}\delta y_{\mathbf{k}\mu}(\tau)=0,
\end{equation}
where $\delta y_{\mathbf{k}\beta}(\tau)$ denotes now the set $\left\{\rho_{\mathbf{k}}, w_{\mathbf{k}\parallel}, \theta_{\mathbf{k}}\right\}$ and $\mathsf{M}$ is the square matrix
\begin{equation}
\label{6.18}
\mathsf{M}=
\left(
\begin{array}{ccc}
0 & ik & 0  \\
ik(1-\sqrt{2}\frac{d}{d+2}\mu\gamma^*\overline{\mu}X) & 2\xi^*-\Theta(k) &ik(1-\sqrt{2}\frac{d}{d+2}\mu\gamma^*D_T^*X)  \\
2\sqrt{2}\zeta^*+\overline{\mu}^*k^2 &ik\left(\frac{2}{d}-2 \kappa_U^*\right) & 2\sqrt{2}\left(\overline{\zeta}_g\gamma^*+\frac{1}{2}\zeta^*\right)+D_{\text{T}}^*k^2 \\
\end{array}
\right),
\end{equation}
where
\beq
\label{6.19}
\Theta(k)\equiv \frac{2\sqrt{2}d}{d+2}\mu\gamma^*\kappa_{U}X-\frac{d-1}{d}\eta^* k^2.
\eeq

\begin{figure}
\begin{center}
\includegraphics[width=.55\columnwidth]{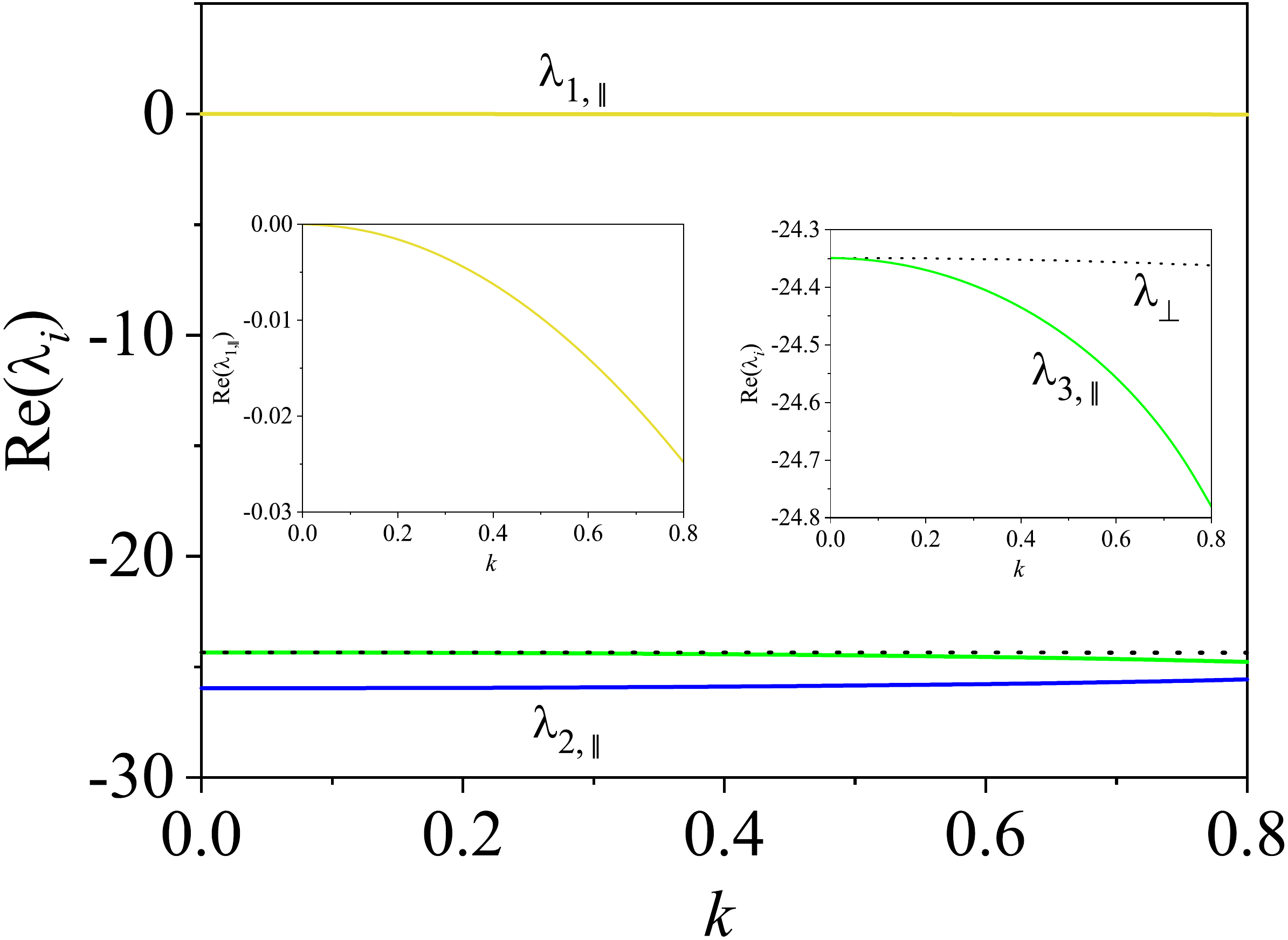}
\end{center}
\caption{Dispersion relations for a three-dimensional granular gas with $\phi=0.001$, $T_g^*=1000$, $m/m_g=1$, and $\alpha=0.8$. From top to bottom the curves correspond to the longitudinal mode $\lambda_{1,||}$, the two degenerate shear (transversal) modes $\lambda_\perp$ (dotted line) and the two remaining longitudinal modes $\lambda_{3,||}$ and $\lambda_{2,||}$. The dependence of $\lambda_{1,||}$, $\lambda_{3,||}$, and $\lambda_\perp$ on $k$ is shown more clearly in the insets. Only the real parts of the eigenvalues are plotted.
\label{fig8}}
\end{figure}

The longitudinal three modes have the form $\text{exp}\left[\lambda_{\ell}(k)\tau\right]$ for $\ell=1,2,3,$. Here, the eigenvalues $\lambda_{\ell}(k)$ of the matrix $\mathsf{M}$ are the solutions of the cubic equation
\begin{equation}
\label{6.20}
\lambda^3+W(k)\lambda^2+Y(k)\lambda+Z(k)=0,
\end{equation}
where
\begin{equation}
\label{6.21}
W(k)=\sqrt{2}\left(\zeta^*+2\overline{\zeta}_g\gamma^*+\sqrt{2}\xi^*\right)-\frac{2\sqrt{2}d}{d+2}\mu\gamma^*\kappa_{U}X
+k^2\left(D_{\text{T}}^*+\frac{d-1}{d}\eta^*\right),
\end{equation}
\beqa
\label{6.22}
Y(k)&=&\frac{d-1}{d}\eta^* D_T^* k^4+k^2 \Bigg\{\frac{2}{d+2}+\sqrt{2}\frac{d-1}{d}\left(\zeta^*+2\overline{\zeta}_g\gamma^*\right)\eta^*-\frac{d}{d+2}\left(\sqrt{2}\mu X \overline{\mu}^*\gamma^*-1\right)\nonumber\\
& & +\frac{2}{d(d+2)}\left(d \kappa_U^*-1\right)\left[d\left(\sqrt{2}\mu X D_T^* \gamma^*-1\right)-2\right]+\frac{2}{d+2}D_T^*\left[(d+2)\xi^*\right.\nonumber\\
& & \left.-\sqrt{2}d\mu X \kappa_U^*\gamma^*\right]\Bigg\}+\frac{2\sqrt{2}}{d+2}\left(\zeta^*+2\overline{\zeta}_g\gamma^*\right)\Big[(d+2)\xi^*-\sqrt{2}d\mu X \kappa_U^*\gamma^*\Big],
\eeqa
\begin{equation}
\label{6.23}
Z(k)=k^2\Bigg\{\sqrt{2}\left(2\overline{\zeta}_g\gamma^*-\zeta^*\right)+\frac{2d}{d+2}\mu X \gamma^*\left[2\left(\zeta^*D_T^*-\overline{\zeta}_g\overline{\mu}^*\gamma^*\right)-\zeta^*\overline{\mu}^*\right]+
k^2\left(D_T^*-\overline{\mu}^*\right)\Bigg\}.
\end{equation}
In the Brownian limit, Eqs.\ \eqref{6.20}--\eqref{6.23} are consistent with those obtained by \cite{GGG19a}. \footnote{There is a typo in Eq.\ (99) of \cite{GGG19a} since the first term of $W(k)$ should be $\sqrt{2}\left(\zeta^*+4\chi^{-1}\gamma^*+2\gamma^*\right)$.}

One of the longitudinal modes $\lambda_{\ell}(k)$ could be unstable for values of the wave number $k<k_{\text{h}}$, where $k_{\text{h}}$ is obtained from Eq.\ \eqref{6.20} when $\lambda=0$, or equivalently, $Z(k_{\text{h}})=0$. This yields the following expressions for $k_{\text{h}}$:
\begin{equation}
\label{6.24}
k_{\text{h}}^{2}=\frac{\sqrt{2}(2\overline{\zeta}_g\gamma^*-\zeta^*)+\frac{2d}{d+2}\mu X \gamma^* \left[2\left(\zeta^*D_T^*-\overline{\zeta}_g\overline{\mu}^*\gamma^*\right)-\zeta^*\overline{\mu}^*\right]}{\overline{\mu}^*-D_T^*}.
\end{equation}
As in the case of $\lambda_\perp(k)$, an study of the dependence of $k_{\text{h}}$ on the parameters of the system shows that $k_{\text{h}}^{2}$ is always negative. Consequently, there are no physical values of the wave number for which the longitudinal modes become unstable and hence, the longitudinal modes are also linearly \emph{stable}.

In summary, the linear stability analysis of the HSS carried out here for a dilute granular gas surrounded by a molecular gas shows no surprises relative to the earlier study performed in the Brownian limit ($m/m_g\to \infty$): the HSS is linearly stable for arbitrary values of the mass ratio $/m_g$. However, the dispersion relations defining the dependence of the eigenvalues $\lambda_\perp(k)$ and $\lambda_{||}(k)$ on the parameter space are very different to those previously obtained when $m/m_g\to \infty$ \cite[]{GGG19a}. As an illustration, Fig.\ \ref{fig8} shows the real parts of the eigenvalues $\lambda_{i,||} (i=1,2,3)$ and $\lambda_{\perp}$ as a function of the wave number $k$ for $\phi=0.001$, $T_g^*=1000$, $m/m_g=1$, and $\alpha=0.8$. It is quite apparent that all the eigenvalues are negative, as expected. In particular, although the longitudinal mode $\lambda_{1,||}$ is quite close to 0, the inset clearly shows that it is always negative. In addition, we also observe that in general the eigenvalues exhibit a very weak dependence on $k$; this contrasts with the results obtained for dry granular fluids [see for instance, Fig.\ 4.7 of \cite{G19}].

\section{Summary and concluding remarks}
\label{sec7}

The main goal of this paper has been to determine the Navier--Stokes--Fourier transport coefficients of a granular gas (modelled as a gas of \emph{inelastic} hard spheres) immersed in a bath of \emph{elastic} hard spheres. We are interested in a situation where the solid particles are sufficiently dilute and hence, one can assume that the state of the molecular gas (bath) is not affected by the presence of grains. Under these conditions, the molecular gas can be considered as a \emph{thermostat} kept at equilibrium at a temperature $T_g$. This system (granular gas thermostated by a molecular gas) was originally proposed years ago by \cite{BMP02a} and it can be considered as a kinetic model for particle-laden suspensions. Thus, in contrast to the previous suspension models employed in the granular literature \cite[]{TK95,SMTK96,GTSH12,SA17} where the effect of the interstitial fluid on grains is accounted for via an effective fluid-solid force, the model considered here takes into account not only the inelastic collisions among grains themselves but also the elastic collisions between particles of the granular and molecular gas. Moreover, we also assume that the volume fraction occupied by the suspended solid particles is very small (low-density regime). In this case, the one-particle velocity distribution function $f(\mathbf{r}, \mathbf{v};t)$ of grains verifies the Boltzmann kinetic equation.

Before analysing inhomogeneous states, we have considered first homogeneous situations. The study of this state is important because it plays the role of the reference base state in the Chapman--Enskog solution to the Boltzmann equation. In this simple situation, the solid particles are subjected to two competing effects. On the one hand, they collide inelastically so that the granular temperature decreases in time. On the other hand, there is an injection of kinetic energy into the system due to their elastic collisions with the more rapid particles of the molecular gas; this effect tends to thermalise the granular gas to the bath temperature $T_g$. In the steady state, both competing effects cancel each other and a breakdown of energy equipartition appears ($T<T_g$). In the HSS, the relevant nonequilibrium parameters are the temperature ratio $T/T_g$ and the kurtosis $a_2$. This latter quantity measures the deviation of the distribution function from its Gaussian (or Maxwellian) form. Both quantities ($T/T_g$ and $a_2$) have been here estimated by considering the so-called first Sonine approximation \eqref{2.4} to the distribution function $f(\mathbf{v})$. In this approximation, the temperature ratio is obtained by numerically solving Eq.\ \eqref{2.2} while the kurtosis is given by Eq.\ \eqref{2.23}. These equations provide the dependence of $T/T_g$ and $a_2$ on the parameter space of the system: the mass ratio $m/m_g$, the (reduced) bath temperature $T_g^*$ [defined by Eq.\ \eqref{2.12}], the volume fraction $\phi$ [defined by Eq.\ \eqref{2.13}], and the coefficient of restitution $\al$. Our theoretical results extend to arbitrary dimensions the results obtained by \cite{S03a} for hard spheres ($d=3$). To assess the accuracy of the (approximate) analytical results, a suite of Monte Carlo simulations have been also performed. Comparison between theory and simulations shows in general a very good agreement, specially in the case of the temperature ratio.

Once the homogeneous state is characterises, the next step has been to solve the Boltzmann equation by means of the Chapman--Enskog--like expansion \cite[]{CC70,BP04,G19}. A subtle point in the expansion is that for small but \emph{arbitrary} perturbations of the HSS, it is expected that the density $n$ and temperature $T$ are specified separately in the \emph{local} reference base state $f^{(0)}(\mathbf{r}, \mathbf{v};t)$ (zeroth-order approximation). This necessarily implies that the temperature is in general a time-dependent parameter (i.e., $\partial_t^{(0)}T\neq 0$). As mentioned in previous works \cite[]{KG13,GGG19a}, this is quite an intricate problem since the complete determination of the transport coefficients in the time-dependent problem requires to solve numerically a set of coupled differential equations. On the other hand, as our main goal is to get the stress tensor and the heat flux to first order in the deviations from the HSS, the corresponding Navier--Stokes--Fourier transport coefficients can be computed to zeroth-order in the deviations from the HSS (steady-state conditions). This simplification allows us to achieve explicit expressions for the above transport coefficients. Their forms are given by Eq.\ \eqref{4.4} for the shear viscosity $\eta$, Eq.\ \eqref{4.9} for the thermal conductivity $\kappa$, Eq.\ \eqref{4.16} for the diffusive heat conductivity $\overline{\mu}$, and Eq.\ \eqref{4.17} for the so-called velocity conductivity coefficient $\kappa_U$. It is quite apparent that the expressions for the scaled coefficients $\eta(\al)/\eta(1)$, $\kappa(\al)/\kappa(1)$,  $n\overline{\mu}(\al)/T\kappa(1)$, and $\kappa_U(\al)/\kappa_U(1)$ [$\eta(1)$, $\kappa(1)$, and $\kappa_U(1)$ being the values of these coefficients for elastic collisions] show a complex dependence on $\al$, $m/m_g$, $T_g^*$, and $\phi$. The dependence on the latter two parameters appears via the (reduced) friction coefficient $\gamma^*$; this coefficient provides a characteristic rate for the elastic collisions between granular and bath particles.

Interestingly, in the Brownian limit ($m/m_g\to \infty$), a careful analysis shows that the expression of $\eta$, $\kappa$, and $\overline{\mu}$ reduce to those previously derived by \cite{GGG19a} by using the Langevin-like model \eqref{1.17} for the instantaneous gas-solid force. In this limiting case, the coefficient $\kappa_U$ vanishes. Therefore, the results reported in this paper extend to arbitrary values of the mass ratio $m/m_g$ the resulting transport coefficients for the particle phase derived in previous works \cite[]{GTSH12,GGG19a}.

As expected, we find that in general the impact of the gas phase on the Navier--Stokes--Fourier transport coefficients is non-negligible. In particular, while the (scaled) shear viscosity coefficient $\eta(\al)/\eta(1)$ increases with decreasing the coefficient of restitution $\al$ for dry (no gas phase) granular gases, Fig.\ \ref{fig4} shows clearly that $\eta(\al)$ is smaller than $\eta(1)$ in the case of granular suspensions. On the other hand, although a more qualitative agreement on the $\al$-dependence of the (scaled) thermal conductivity coefficient $\kappa(\al)/\kappa(1)$ is found for dry granular gases and gas-solid suspensions, significant quantitative differences between both systems appear as the inelasticity in collisions increases. Finally, the differences in the case of the diffusive heat conductivity $\overline{\mu}$ are much more important since the magnitude of the (scaled) coefficient $n\overline{\mu}/T\kappa(1)$ of (dry) granular gases is much more large than that of granular suspensions [see Fig.\ \ref{fig6}].

The knowledge of the forms of the transport coefficients opens up the possibility of preforming a linear stability analysis on the resulting continuum hydrodynamic equations. As in the Brownian limit case \cite[]{GGG19a}, the analysis shows that the HSS is always linearly stable whatever the mass ratio considered is.

One of the main limitations of the results derived in this paper is its restriction to the low-density regime. The extension of the present theory to a moderately dense gas-solid suspension described by the Enskog kinetic equation is an interesting project for the future. These results could stimulate the performance of MD simulations to asses the reliability of the theory for finite densities. Another challenging work could be the determination of the non-Newtonian rheological properties of a granular suspension under simple shear flow. This study would allow to extend previous studies \cite[]{TK95,SMTK96,ChVG15,SA17,ASG19,THSG20} to arbitrary values of the mass ratio $m/m_g$. Another possible project could be to revisit the results obtained in this paper by considering the charge transport equation recently considered by \cite{CKO21}. Work along these lines will be carried out in the near future.\\

 \noindent \textbf{Acknowledgements.} The authors acknowledge financial support from Grant PID2020-112936GB-I00 funded by MCIN/AEI/ 10.13039/501100011033, and from Grants IB20079 and GR18079 funded by Junta de Extremadura (Spain) and by ERDF A way of making Europe. The research of R.G.G. also has been supported by the predoctoral fellowship BES-2017-079725 from the Spanish Government.\\

\noindent\textbf{Declaration of interests.} The authors report no conflict of interest.\\

\noindent\textbf{Author ORCIDs.}
\\ Rubén Gómez González: https://orcid.org/0000-0002-5906-5031
\\Vicente Garzó: https://orcid.org/0000-0001-6531-9328


\appendix
\section{First-order distribution function}
\label{appA}

Some technical details employed in the derivation of the kinetic equation for the first-order distribution $f^{(1)}$ are given in this appendix. To first order in the spatial gradients, the velocity distribution function $f^{(1)}(\mathbf{r}, \mathbf{v};t)$ verifies the Boltzmann kinetic equation
\beq
\label{a1}
\partial_t^{(0)}f^{(1)}+\mathcal{L}f^{(1)}-J_g[f^{(1)},f_g^{(0)}]=-\Big(D_t^{(1)}+\mathbf{V}\cdot \nabla \Big)f^{(0)}-\frac{m_g}{T_g}\Delta \mathbf{U}\cdot J_g[f^{(0)},\mathbf{V}f_g^{(0)}],
\eeq
where $D_t^{(1)}\equiv \partial_t^{(1)}+\mathbf{U}\cdot \nabla$ and the linear operator $\mathcal{L}$ is defined by Eq.\ \eqref{3.18}. To first order, the macroscopic balance equations \eqref{1.8}--\eqref{1.10} are
\beq
\label{a2}
D_t^{(1)}n=-n \nabla \cdot \mathbf{U}, \quad D_t^{(1)}T=-\frac{2p}{dn}\nabla\cdot \mathbf{U}-T\left(\zeta^{(1)}+\zeta_g^{(1)}\right),
\eeq
\beq
\label{a3}
D_t^{(1)}\mathbf{U}=-\rho^{-1}\nabla p-\xi \Delta \mathbf{U}+\rho^{-1}\boldsymbol{\mathcal{K}}[f^{(1)}],
\eeq
where $\zeta^{(1)}$ and $\zeta_g^{(1)}$ are the first-order contributions to the production rates, the operator $\boldsymbol{\mathcal{K}}[X]$ is defined by Eq.\ \eqref{3.19}, and $\xi$ is defined in Eq.\ \eqref{3.25.1}. The production rates are functional of the distribution $f^{(1)}$ and their explicit forms are given by Eqs.\ \eqref{3.26}--\eqref{3.28}.

The use of the balance equations \eqref{a2} and \eqref{a3} allows one to compute the first term on the right side of Eq.\ \eqref{a1}. The result is
\beqa
\label{a4}
-\Big(D_t^{(1)}+\mathbf{V}\cdot \nabla \Big)f^{(0)}&=&\mathbf{A}\cdot \nabla \ln T+\mathbf{B}\cdot \nabla \ln n+C_{ij}\frac{1}{2}
\Big(\partial_i U_j+\partial_j U_i-\frac{2}{d}\delta_{ij}\nabla \cdot \mathbf{U}\Big)\nonumber\\
& & +D\nabla\cdot \mathbf{U}+\mathbf{E}\cdot \Delta \mathbf{U}+\rho^{-1}\frac{\partial f^{(0)}}{\partial \mathbf{V}}\cdot \boldsymbol{\mathcal{K}}[f^{(1)}]\nonumber\\
& & +T\left(\zeta^{(1)}+\zeta_g^{(1)}\right)\frac{\partial f^{(0)}}{\partial T},
\eeqa
where $\partial_i\equiv \partial/\partial r_i$ and the quantities $\mathbf{A}$, $\mathbf{B}$, $C_{ij}$, $D$, and $\mathbf{E}$ are given by Eqs.\ \eqref{3.20}--\eqref{3.25}, respectively. Substitution of Eq.\ \eqref{a4} into Eq.\ \eqref{a1} yields
\beqa
\label{a5}
& & \partial_t^{(0)}f^{(1)}-T\left(\zeta^{(1)}+\zeta_g^{(1)}\right)\frac{\partial f^{(0)}}{\partial T}+\mathcal{L}f^{(1)}-J_g[f^{(1)},f_g^{(0)}]-
\rho^{-1}\frac{\partial f^{(0)}}{\partial \mathbf{V}}\cdot \boldsymbol{\mathcal{K}}[f^{(1)}]\nonumber\\
& & =\mathbf{A}\cdot \nabla \ln T+\mathbf{B}\cdot \nabla \ln n+C_{ij}\frac{1}{2}
\Big(\partial_i U_j+\partial_j U_i-\frac{2}{d}\delta_{ij}\nabla \cdot \mathbf{U}\Big)+D\nabla\cdot \mathbf{U}\nonumber\\
& & +\Big(\mathbf{E}-\frac{m_g}{T_g}J_g[f^{(0)},\mathbf{V}f_g^{(0)}]\Big)\cdot \Delta \mathbf{U}.
\eeqa
The solution of Eq.\ \eqref{a5} is of the form
\beqa
\label{a6}
f^{(1)}(\mathbf{V})&=&\boldsymbol{\mathcal{A}}(\mathbf{V})\cdot \nabla \ln T+\boldsymbol{\mathcal{B}}(\mathbf{V})\cdot \nabla \ln n+\mathcal{C}_{ij}\frac{1}{2}
\Big(\partial_i U_j+\partial_j U_i-\frac{2}{d}\delta_{ij}\nabla \cdot \mathbf{U}\Big)\nonumber\\
& & +\mathcal{D}(\mathbf{V}) \nabla \cdot \mathbf{U}+\boldsymbol{\mathcal{E}}\cdot \Delta \mathbf{U}.
\eeqa
Substitution of this into Eq.\ \eqref{a5} gives the integral equations \eqref{3.13}--\eqref{3.17}. Upon obtaining these equations, we have taken into account that
\beq
\label{a7}
\partial_t^{(0)}\left\{\boldsymbol{\mathcal{A}}, \boldsymbol{\mathcal{B}}, \mathcal{C}_{ij}, \mathcal{D}, \boldsymbol{\mathcal{E}}\right\}=-\left(\zeta^{(0)}+\zeta_g^{(0)}\right)T\partial_T \left\{\boldsymbol{\mathcal{A}}, \boldsymbol{\mathcal{B}}, \mathcal{C}_{ij}, \mathcal{D}, \boldsymbol{\mathcal{E}}\right\},
\eeq
and the result
\beqa
\label{a8}
\partial_t^{(0)}\nabla \ln T=\nabla \partial_t^{(0)} \ln T&=&-\nabla \left(\zeta^{(0)}+\zeta_g^{(0)}\right)=-\Bigg[\zeta^{(0)}+\zeta_g^{(0)}\Bigg(1-\varepsilon \frac{\partial \ln \zeta_g^*}{\partial \varepsilon}\Bigg)\Bigg]\nabla \ln n\nonumber\\
& &-
\frac{1}{2}\Bigg[\zeta^{(0)}+\zeta_g^{(0)}\Bigg(1+2\chi \frac{\partial \ln \zeta_g^*}{\partial \chi}\Bigg)\Bigg]\nabla \ln T.
\eeqa
Here, $\varepsilon$ is defined in Eq.\ \eqref{2.12} and $\chi\equiv T/T_g$.

\section{Leading Sonine approximations to the Navier--Stokes--Fourier transport coefficients}
\label{appB}

In this appendix, we determine the explicit expressions of the Navier--Stokes--Fourier transport coefficients in steady-state conditions. To obtain them, we consider the leading Sonine approximations to the unknowns $\boldsymbol{\mathcal{A}}$, $\boldsymbol{\mathcal{B}}$, $\mathcal{C}_{ij}$, $\mathcal{D}$, and $\boldsymbol{\mathcal{E}}$ and neglect non-Gaussian corrections to the zeroth-order distribution $f^{(0)}$ (i.e., we take $a_2=0$). Since the procedure to obtain these expressions is quite similar to the one employed in some previous works on granular binary mixtures \cite[]{GD02,GM07}, only some partial results will be displayed in this appendix.

\subsection{Leading Sonine approximation to $\eta$}

In the case of the shear viscosity $\eta$, the leading Sonine approximation to $\mathcal{C}_{ij}(\mathbf{V})$ (lowest degree polynomial) is
\beq
\label{b1}
\mathcal{C}_{ij}(\mathbf{V})\to -f^{(0)}(\mathbf{V}) R_{ij}(\mathbf{V}) \frac{\eta}{n T^2},
\eeq
where $f^{(0)}$ is the Maxwellian distribution \eqref{4.1} of the granular gas, the polynomial $R_{ij}$ is given by Eq.\ \eqref{3.33.2}, and $\eta$ is defined in Eq.\ \eqref{3.31}. Since $R_{ij}(\mathbf{V})$ is an even polynomial in $\mathbf{V}$, then $\mathcal{K}_{\ell}[R_{ij}]=0$, and the integral equation \eqref{4.0.2} reads
\beq
\label{b2}
-\frac{\eta}{nT^2}\left\{\mathcal{L}[f^{(0)}\mathcal{R}_{ij}]-J_g[f^{(0)}\mathcal{R}_{ij},f_g^{(0)}]\right\}={C}_{ij}.
\eeq
To determine $\eta$, we multiply both sides of Eq.\ \eqref{b2} by $R_{ij}(\mathbf{V})$ and integrate over velocity. The result can be written as
\beq
\label{b3}
\frac{1}{(d-1)(d+2)}\frac{\eta}{nT^2}\Bigg\{\int d\mathbf{v} R_{ij}(\mathbf{V}) \mathcal{L}\left[f^{(0)}R_{ij}\right]-\int d\mathbf{v} R_{ij}(\mathbf{V}) J_g\left[f^{(0)}R_{ij},f_g^{(0)}\right]\Bigg\}=p.
\eeq
The collision integral involving the linearised Boltzmann collision operator $\mathcal{L}$ is given by \cite[]{BDKS98}
\beq
\label{b4}
\frac{1}{(d-1)(d+2)}\frac{1}{nT^2}\int d\mathbf{v} R_{ij}(\mathbf{V}) \mathcal{L}\left[f^{(0)}R_{ij}\right]=\nu_\eta^*\nu_0,
\eeq
where $\nu_\eta^*$ is defined by Eq.\ \eqref{4.7} and $\nu_0=p/\eta_0$, $\eta_0$ being the shear viscosity of a dilute gas of elastic hard spheres [see Eq.\ \eqref{4.5}]. The collision integral involving the Boltzmann--Lorentz operator $J_g$ can be obtained from previous works on granular mixtures \cite[]{GD02,GM07} when one particularises to elastic collisions. In terms of the friction coefficient $\gamma$, the result is
\beq
\label{b5}
-\frac{1}{(d-1)(d+2)}\frac{1}{nT^2}\int d\mathbf{v} R_{ij}(\mathbf{V}) J_g\left[f^{(0)}R_{ij},f_g^{(0)}\right]=\widetilde{\nu}_\eta \gamma,
\eeq
where $\widetilde{\nu}_\eta$ is given by Eq.\ \eqref{4.8}. The expression \eqref{4.4} for $\eta$ can be easily obtained when one takes into account Eqs.\ \eqref{b4} and \eqref{b5} in Eq.\ \eqref{b3}.

\subsection{Leading Sonine approximation to $\kappa$, $\overline{\mu}$, and $\kappa_U$}

The heat flux transport coefficients $\kappa$, $\overline{\mu}$, and $\kappa_U$ are defined by Eqs.\ \eqref{3.32}--\eqref{3.33.1}, respectively. The leading Sonine approximation to $\boldsymbol{\mathcal{A}}$, $\boldsymbol{\mathcal{B}}$, and $\boldsymbol{\mathcal{E}}$ is
\beq
\label{b6}
\boldsymbol{\mathcal{A}}(\mathbf{V})\to c_\kappa \mathbf{S}(\mathbf{V}) f^{(0)}(\mathbf{V}), \quad  \boldsymbol{\mathcal{B}}(\mathbf{V})\to c_\mu \mathbf{S}(\mathbf{V}) f^{(0)}(\mathbf{V}), \quad \boldsymbol{\mathcal{E}}(\mathbf{V})\to c_{\kappa_U} \mathbf{S}(\mathbf{V}) f^{(0)}(\mathbf{V}),
\eeq
where the Sonine coefficients $c_\kappa$, $c_\mu$, and $c_{\kappa_U}$ are defined, respectively, as
\beq
\label{b7}
\left(
\begin{array}{c}
c_\kappa\\
c_\mu \\
c_{\kappa_U}
\end{array}
\right)
=\frac{2}{d(d+2)}\frac{m}{nT^3}\int d\mathbf{V}\;
\left(
\begin{array}{c}
\boldsymbol{\mathcal{A}}(\mathbf{V})\cdot \mathbf{S}(\mathbf{V})\\
\boldsymbol{\mathcal{B}}(\mathbf{V})\cdot \mathbf{S}(\mathbf{V})\\
\boldsymbol{\mathcal{E}}(\mathbf{V})\cdot \mathbf{S}(\mathbf{V})
\end{array}
\right)=-\left(
\begin{array}{c}
\frac{2}{d+2}\frac{m}{nT^2}\kappa\\
\frac{2}{d+2}\frac{m}{T^3}\overline{\mu} \\
\frac{2}{d+2}\frac{m}{nT^3}\kappa_U
\end{array}
\right).
\eeq
Using the Sonine approximations \eqref{b6}, the collision integral \eqref{3.19} [when $X=S_i(\mathbf{V})$] is
\beq
\label{b8}
\mathcal{K}_i[S_j]=\int d\mathbf{v}\; m V_i\;  J_g[S_j f^{(0)},f_g^{(0)}]=-\delta_{ij} \frac{1}{2}n T^2\mu \theta^{-1/2}\left(1+\theta\right)^{-1/2}
\gamma.
\eeq
Taking into account Eq.\ \eqref{b8}, the integral equation \eqref{4.0.1} becomes
\beqa
\label{b9}
& &-\frac{2}{d+2}\frac{m}{nT^2}\kappa\left\{
-\zeta_g^{(0)}\chi \frac{\partial \ln \zeta_g^*}{\partial \chi}f^{(0)}\mathbf{S}+\mathcal{L}[f^{(0)}\mathbf{S}]-J_g[f^{(0)}\mathbf{S},f_g^{(0)}]\right\}=\mathbf{A}\nonumber\\
& & +\frac{1}{d+2}\frac{\mu\gamma }{n}\theta^{-1/2}\left(1+\theta\right)^{-1/2}\frac{\partial f^{(0)}}{\partial \mathbf{V}} \kappa,
\eeqa
where use has been made of the Sonine approximation \eqref{b6} to $\boldsymbol{\mathcal{A}}$. As in the case of the shear viscosity, $\kappa$ is determined by multiplying both sides of Eq.\ \eqref{b9} by $\mathbf{S}(\mathbf{V})$ and integrating over $\mathbf{V}$. After some algebra, one achieves
\beq
\label{b10}
-\zeta_g^{(0)}\chi \frac{\partial \ln \zeta_g^*}{\partial \chi} \kappa + \frac{2}{d(d+2)}\frac{m}{nT^3}\kappa\Bigg\{\int d\mathbf{V} \mathbf{S}\cdot \mathcal{L}\left[f^{(0)}\mathbf{S}\right]-\int d\mathbf{V} \mathbf{S}\cdot J_g\left[f^{(0)}\mathbf{S},f_g^{(0)}\right]\Bigg\}=\frac{d+2}{2}\frac{p}{m},
\eeq
where we have accounted for the results
\beq
\label{b11}
-\frac{1}{d T}\int d\mathbf{V}\; \mathbf{S}(\mathbf{V})\cdot \mathbf{A}(\mathbf{V})=\frac{d+2}{2m}p, \quad \int d\mathbf{V}\; \mathbf{S}(\mathbf{V})\cdot \frac{\partial f^{(0)}}{\partial \mathbf{V}}=0.
\eeq
The corresponding collision integrals can be written as \cite[]{BDKS98,GD02,GM07}
\beq
\label{b12}
\frac{2}{d(d+2)}\frac{m}{nT^3}\int d\mathbf{V} \mathbf{S}\cdot \mathcal{L}\left[f^{(0)}\mathbf{S}\right]=\nu_\kappa^* \nu_0,
\eeq
\beq
\label{b13}
-\frac{2}{d(d+2)}\frac{m}{nT^3}\int d\mathbf{V} \mathbf{S}\cdot J_g\left[f^{(0)}\mathbf{S},f_g^{(0)}\right]=\widetilde{\nu}_\kappa \gamma,
\eeq
where $\nu_\kappa^*$ is defined by Eq.\ \eqref{4.12} while $\widetilde{\nu}_\kappa$ is given by Eqs.\ \eqref{4.13}--\eqref{4.15}.
In addition, according to the expression \eqref{2.7} of $\widetilde{\zeta}_g^{(0)}$, one gets the relation
\beq
\label{b14}
-\zeta_g^{(0)}\chi \frac{\partial \ln \zeta_g^*}{\partial \chi}=\beta \gamma, \quad \beta=\left(x^{-1}-3x\right)\mu^{3/2}\chi^{-1/2}.
\eeq
Substitution of Eqs.\ \eqref{b12}--\eqref{b14} into Eq.\ \eqref{b10} leads to the expression \eqref{4.9} for $\kappa$.

The evaluation of the diffusive heat conductivity $\overline{\mu}$ follows similar steps to those carried out in the evaluation of $\kappa$. Taking into account the leading Sonine approximations \eqref{b6} to $\boldsymbol{\mathcal{A}}$ and $\boldsymbol{\mathcal{B}}$, the integral equation \eqref{4.0.2} reads
\beqa
\label{b15}
& &
-\frac{2}{d+2}\frac{m}{T^3}\overline{\mu}\left\{
\mathcal{L}[f^{(0)}\mathbf{S}]-J_g[f^{(0)}\mathbf{S},f_g^{(0)}]\right\}=\mathbf{B}+\frac{1}{d+2}\frac{\mu\gamma }{T}\theta^{-1/2}\left(1+\theta\right)^{-1/2}\frac{\partial f^{(0)}}{\partial \mathbf{V}} \overline{\mu}\nonumber\\
& & -\frac{2}{d+2}\frac{m}{nT^2}\zeta_g^{(0)}\kappa f^{(0)}\mathbf{S},
\eeqa
where use has been made of Eq.\ \eqref{b8} and the result
\beq
\label{b16}
\zeta_g^{(0)}\varepsilon \frac{\partial \ln \zeta_g^*}{\partial \varepsilon}=\zeta_g^{(0)}.
\eeq
Multiplying both sides of Eq.\ \eqref{b15} by $\mathbf{S}(\mathbf{V})$ and integrating over velocity, one gets
\beq
\label{b17}
\left(\nu_0\nu_\kappa^*+\widetilde{\nu}_\kappa \gamma \right)\overline{\mu}=\frac{T}{n} \zeta^{(0)}\kappa,
\eeq
where use has been made of the steady state condition $\zeta_g^{(0)}=-\zeta^{(0)}$. The solution to Eq.\ \eqref{b17} yields the expression \eqref{4.16} for $\overline{\mu}$.

Finally, we consider the coefficient $\kappa_U$. As said in the main text, it is a new transport coefficient not present for dry granular monocomponent gases. Taking into account the expression \eqref{4.2} of $\xi$, Eq.\ \eqref{b8}, and the leading Sonine approximation \eqref{b6} to $\boldsymbol{\mathcal{E}}$, the integral equation \eqref{4.0.5} reads
\beqa
\label{b18}
& & -\frac{2}{d+2}\frac{m}{nT^3}\kappa_U \left\{
\mathcal{L}[f^{(0)}\mathbf{S}]-J_g[f^{(0)}\mathbf{S},f_g^{(0)}]\right\}=-\mu\;  \theta^{-1/2}(1+\theta)^{1/2}\gamma \frac{\partial f^{(0)}}{\partial \mathbf{V}}\nonumber\\
& & +\frac{1}{d+2}\frac{\mu\gamma }{n T}\theta^{-1/2}\left(1+\theta\right)^{-1/2}\frac{\partial f^{(0)}}{\partial \mathbf{V}}\kappa_U
-\frac{m_g}{T_g} J_g[f^{(0)},\mathbf{V}f_g^{(0)}].
\eeqa
As in the cases of $\kappa$ and $\overline{\mu}$, one multiplies both sides of Eq.\ \eqref{b8} by $\mathbf{S}(\mathbf{V})$ and integrates over velocity to get
\beq
\label{b19}
\left(\nu_0\nu_\kappa^*+\widetilde{\nu}_\kappa \gamma \right)\kappa_U=-\frac{1}{2}n T \mu (1+\theta)^{-1/2}\theta^{-1/2}B \gamma,
\eeq
where use has been made of the result
\beq
\label{b20}
\frac{m_g}{T_g} \int d \mathbf{V} \mathbf{S}\cdot J_g[f^{(0)},\mathbf{V} f_g^{(0)}]=-\frac{d nT}{2}\mu \gamma (1+\theta)^{-1/2}\theta^{-1/2}\; B,
\eeq
where the expression of the quantity $B$ is displayed in Eq.\ \eqref{4.18}. The expression \eqref{4.17} for $\kappa_U$ can be easily obtained from Eq.\ \eqref{b19}.

\bibliographystyle{jfm}

\bibliography{Brownian}

\end{document}